# An entropy-controlled objective chip for reflective confocal microscopy with subdiffraction-limit resolution

Jun He[1,#], Dong Zhao[1,#], Hong Liu[2], Jinghua Teng[2,*], Cheng-Wei Qiu[3,*], Kun Huang[1,*]

[1]Department of Optics and Optical Engineering, University of Science and Technology of China, Hefei, Anhui 230026, China

[2]Institute of Materials Research and Engineering, Agency for Science Technology and Research (A*STAR), 2 Fusionopolis Way, #08-03, Innovis, Singapore 138634, Singapore

[3]Department of Electrical and Computer Engineering, National University of Singapore, 4 Engineering Drive 3, Singapore 117576, Singapore

[#] *J. H. and D. Z.* contributed equally to this work.

[*]Corresponding authors: K. H. (huangk17@ustc.edu.cn), J. T. (jh-teng@imre.a-star.edu.sg) or C. Q. (chengwei.qiu@nus.edu.sg)

## Abstract

**Planar lenses with optimized but disordered structures can focus light beyond the diffraction limit. However, these disordered structures have inevitably destroyed wide-field imaging capability, limiting their applications in microscopy. Here we introduce information entropy $S$ to evaluate the disorder of an objective chip by using the probability of its structural deviation from standard Fresnel zone plates. Inspired by the theory of entropy change, we predict an equilibrium point $S_0=0.5$ to balance wide-field imaging (theoretically evaluated by the Strehl ratio) and subdiffraction-limit focusing. To verify this, a $NA=0.9$ objective chip with a record-long focal length of 1 mm is designed with $S=0.535$, which is the nearest to the equilibrium point among all reported planar lenses. Consequently, our fabricated chip can focus light with subdiffraction-limit size of $0.44\lambda$ and image fine details with spatial frequencies up to 4000 lp/mm in experiment. These unprecedented performances enable ultracompact reflective confocal microscopy for superresolution imaging.**

## Introduction

Since the first microscope was invented in 1595 by a Dutch father-son team Hans and Zacharias Janssen[1], optical objectives have developed rapidly with improved performance in aberration correction, field of view, magnification, and numerical aperture, except its imaging resolution being limited by diffraction of light[2]. Based on refractive optics[3], traditional objectives need multiple elements with carefully designed curvatures and air intervals for better imaging, thus leading to a bulky volume. The situation of objectives has been sustained for over 400 years until subwavelength-thick metalenses were reported with ultracompact volume in 2016[4]. However, high-aspect-ratio dielectric nanobricks in the metalenses exhibit high efficiency under normal incidence only with a small tolerance of tilting angle[5], resulting in large spatial-frequency details of objects being rejected by the metalenses operating in an imaging mode[6, 7]. Consequently, the imaging resolutions of the metalenses and their related scanning confocal microscopies (SCMs) are diffraction-limited to $0.51\lambda/NA$ ($\lambda$ is the wavelength of light and $NA$ is the numerical aperture of the metalens)[8].

Efforts to enhance the resolving power of objectives have also been made by using planar superoscillation lenses[9-12] and supercritical lenses[13-15] that realize subwavelength focal spots by optimizing the destructive or constructive interference between multiple diffracting beams[7, 16-18]. Since only the focusing properties of these planar lenses are designed without considering the capability of direct wide-field imaging, superoscillation and supercritical lenses are usually used as optical probes in an SCM[7], where an additional refraction-based objective is mandatorily required to collect the transmitted light through the objects. This leads to the fact that all these planar-lens-based SCMs must operate in transmission mode and are only valid for objects sitting on a transparent substrate that introduces spherical aberration, which requires collection objectives with coverslip collars for correction[2]. Without involving these issues, reflective SCMs are, therefore, more popular for noninvasive and in vivo imaging of various specimens[19]. Despite the strong requirements from applications, it is still difficult to demonstrate reflective planar-lens-based SCMs with better resolving power than commercial SCMs, due to the lack of high-performance planar objectives that possess dual functionalities of focusing and imaging beyond the diffraction limit.

The challenges to desigining such a planar objective are twofold. First, designing a planar objective for subdiffraction-limit focusing generally leads to an irregularly distributed phase or amplitude[18]. However, such structural disorder is not preferred in its imaging counterpart, where the analytical phase or amplitude is preferred for the constructive reconstruction of objects[7]. Such a dilemma is a fundamental barrier to demonstrating planar objectives with subdiffraction-limit resolution in both focusing and imaging. Second, the diameter of the planar objectives should be sufficiently large to suppress the diffraction effect of collected light by the planar objectives (when working in the imaging mode) for better image formation. Correspondingly, the number of fine structures in the planar objectives is extremely large because of the short wavelength in the visible spectrum, thereby increasing the technical difficulty in the design and fabrication of planar objectives.

To overcome these challenges, we propose a disorder-controlled objective chip that functionally integrates a binary-phase Fresnel zone plate (FZP) and a weakly perturbed few-ring phase mask into a single ultrathin element. By introducing the concept of information entropy, we theoretically predict that an objective chip with its entropy at $S_0=0.5$ can maintain the imaging and superfocusing properties simultaneously. Using deep-ultraviolet (DUV) lithography, the fabricated objective chip experimentally exhibits a focal spot of $0.44\lambda$ (below the Rayleigh criterion of $0.51\lambda/NA=0.57\lambda$) without strong sidebands and the capability of imaging fine objects with spatial frequency of 4000 lp/mm. Benefiting from this, an ultracompact reflective SCM is built with an imaging resolution (center-to-center) of 200 nm at $\lambda=405$ nm and a record-long working distance of 1 mm, superior to the state-of-the-art SCMs.

**Disorder of objective chip**

Since standard FZPs have analytical complex transmission, *i.e.*, $U_{FZP}=A_{FZP}e^{i\varphi_{FZP}}$ (where $A_{FZP}$ and $\varphi_{FZP}$ are the amplitude and phase modulation, respectively), they can realize wide-field but diffraction-limit imaging. For an objective chip, its complex transmission $U_{chip}=A_{PL}e^{i\varphi_{PL}}$ deviates irregularly from that of an FZP, thus creating undesired disorder for imaging purposes. Assuming that the deviations are $\Delta A=A_{chip}-A_{FZP}$ for amplitude and $\Delta\varphi=\varphi_{chip}-\varphi_{FZP}$ for phase, we can rewrite the transmission of the objective chip as $U_{chip}=(A_{FZP}+\Delta A)e^{i(\varphi_{FZP}+\Delta\varphi)}=A_{FZP}e^{i\varphi_{FZP}}(1+\Delta A/A_{FZP})e^{i\Delta\varphi}=U_{FZP}\cdot U_\Delta$, where the deviated transmission $U_\Delta=(1+\Delta A/A_{FZP})e^{i\Delta\varphi}$ introduces optical aberration in imaging but offers more degrees of freedom for subdiffraction-limit focusing. Thus, for an arbitrary objective chip, its complex transmission contains a standard imaging part $U_{FZP}$ (*i.e.*, FZP) and an additional aberration part $U_\Delta$. In the design of an objective chip, the item $U_\Delta$ is fundamentally important in building the connection between imaging and focusing.

Considering that $U_\Delta$ is distributed spatially in an irregular way, we first investigate its statistical property of such deviation by defining a dimensionless parameter—deviation probability

$$p_1 = \sum_{m=0}^{M-1} \frac{1}{M} \left| \frac{\int_{r_m}^{r_{m+1}} \Delta \cdot r \, dr}{\int_{r_m}^{r_{m+1}} \Delta_{max} \cdot r \, dr} \right|, \tag{1}$$

where $\Delta = \Delta A$ with a maximum value of $\Delta_{max} = 1$ for the amplitude deviation, $\Delta = \Delta\varphi$ with a maximum value of $\Delta_{max} = \pi$ for the phase deviation, the objective chip is divided into minimum diffraction subunits (*i.e.*, zones in the FZP) with their boundaries of $r_m = \sqrt{m\lambda f + (m\lambda)^2/4}$ (*m* is the index of the *m*-th zone, λ is the wavelength and *f* is the focal length of the lens), *M* is the total number of zones contained in the FZP, and the modulus is used to ensure the non-negative probability. Because the standard FZPs have pure amplitude (*i.e.*, $\varphi_{FZP} = 0$ in $U_{FZP}$ for binary-amplitude FZPs) or phase (*i.e.*, $A_{FZP}$ is a constant in $U_{FZP}$ for binary-phase FZPs) modulation, we only need to calculate *ΔA* or *Δφ* in Eq. (1) for most objective chips. According to its definition, the deviation probability *p*₁ ranges from 0 to 1, leaving a probability *p*₂=1-*p*₁ for the unchanged part. It behaves like an information channel with binary values, where the entropy $S = -\sum_{i=1}^{2} p_i \log_2 p_i$ is usually used to evaluate the disorder of this information system[20].

Similarly, based on our defined probability *p*₁ and *p*₂, we can also calculate the disorder of an objective chip by using the information entropy *S*. When *p*₁=0 or *p*₁=1, the corresponding *S* equals zero, which means high certainty without any disorder. It agrees with the real cases that both objective chips with *p*₁=0 and *p*₁=1 refer to standard FZPs, where $U_\Delta = 1$. When *p*₁=0.5, the entropy *S*=1, which implies the highest disorder because half of the zones are reversed randomly. Although high disorder offers large degrees of freedom for optical super-focusing, it also destroys the imaging capability of the objective chip due to optical aberration dominated by its random $U_\Delta$. Thus, we infer that the entropy *S* is symmetric about *p*₁=0.5, where the peak is located. At the side of 0≤ *p*₁ ≤0.5, the entropy *S* increases monotonically from 0 to 1. High certainty at *S*=0 is helpful in wide-field imaging but high disorder at *S*=1 is required to realize super-focusing.

To obtain a good balance between imaging and super-focusing, we introduce a thermodynamic analog that the change of entropy for an isolated system originates from outside work or heat transfer[21]. In our case, a virtual work *W* is assumed to govern the change *ΔS* of entropy by following a straightforward relationship $W \propto \Delta S$. The change of entropy from *S*=0 to *S*₀ requires the virtual work *W*₁, while the change from *S*₀ to *S*=1 requires virtual work *W*₂. We suggest that, when both virtual works are equal, *i.e.*, *W*₁=*W*₂, good balance between imaging and super-focusing is achieved with an equality $\Delta S_1 = \Delta S_2$ (*i.e.*, $S_0 - 0 = 1 - S_0$), leading to the equilibrium point *S*₀=0.5.

Interestingly, at this equilibrium point, its relative deviation probability $p_1$ equals 0.11, which is smaller than the middle point $p_1=0.25$ of the interested range $0 \leq p_1 \leq 0.5$. This implies that the entropy $S$ is more sensitive to the intrinsic disorder of the objective chip than the deviation probability $p_1$, thereby indicating the rationality of the proposed equilibrium point $S_0=0.5$.

**Strehl ratio and focal size of the objective chip**

To quantitatively investigate the imaging and focusing properties of an objective chip with different disorders, a binary-phase objective chip (Fig. 1a) with a focal length $f=1$mm and $NA=0.9$ (implying its diameter of 4.13 mm) is exemplified here to enhance the optical efficiency in both focusing and imaging. Compared with its corresponding binary-phase FZP, this proposed objective chip has only the phase deviation of $\Delta\varphi$ because of $\Delta A=0$, leaving $U_\Delta = e^{i\Delta\varphi}$. This implies that any binary-phase objective chip can be taken as a combination of an analytical FZP and an additional few-ring phase (*i.e.*, $\Delta\varphi$) mask, as sketched in Fig. 1b. Since the phase $e^{i\Delta\varphi}$ in the few-ring mask introduces optical aberration, we evaluate its influence on imaging quality by using the relative Strehl ratio (SR)[3]

$$SR = \frac{I_{chip}(0,0,z=f)}{I_{FZP}(0,0,z=f)} = \frac{\left|\iint E_0(r,\varphi)\cdot e^{i(\varphi_{FZP}+\Delta\varphi)}\frac{exp(ikR)}{R^3}(ikR-1)zrdrd\varphi\right|^2}{\left|\iint E_0(r,\varphi)\cdot e^{i\varphi_{FZP}}\frac{exp(ikR)}{R^3}(ikR-1)zrdrd\varphi\right|^2}, \quad (2)$$

where the incident electric field $E_0$ is taken as unity in this work, the wavenumber $k=2\pi/\lambda$, $R^2=(u-x)^2+(v-y)^2+z^2$, $r^2=x^2+y^2$, $tan\varphi=y/x$, $x$ and $y$ are the Cartesian coordinates at the initial plane of the objective chip, $u=v=0$ and $z=f$ are used to obtain the on-axis intensity at the focal plane, and $I_{FZP}$ is fixed once the FZP is given. In Eq. (2), the on-axis intensity of the FZP, having the same modulation (phase or amplitude) type as that of the designed objective chip, is used as the denominator to avoid the influence of the focusing efficiency of the FZP with different modulation types. Therefore, Eq. (2) defines a relative Strehl ratio, which is more useful in evaluating the optical aberration of imaging systems.

For a binary-phase objective chip with its deviation probability $p_1$ (only $0 \leq p_1 \leq 0.5$ is considered in the following because the entropy $S$ is symmetric about $p_1=0.5$), its relative Strehl ratio can be approximated as $SR = \left(1 - 2a_0 p_1 M/\sqrt{I_{FZP}}\right)^2$, where the on-axis intensity $a_0$ of each zone plate ranges from $a_{0min}=0.87$ to $a_{0max}=2$, see the detailed derivations in Supplementary Section 1. Thus, for a given $p_1$, we can analytically obtain the range of SR: 1) $SR_{min} =$

$(1 - 2a_{0max}p_1 M/\sqrt{I_{FZP}})^2$ for $0 \leq p_1 \leq \sqrt{I_{FZP}}/(2Ma_{0max})$ and $SR_{min} = 0$ for $\sqrt{I_{FZP}}/(2Ma_{0max}) \leq p_1 \leq 0.5$ ; 2) $SR_{max} = (1 - 2a_{0min}p_1 M/\sqrt{I_{FZP}})^2$ for $0 \leq p_1 \leq \sqrt{I_{FZP}}/(Ma_{0max} + Ma_{0min})$ and $SR_{max} = (1 - 2a_{0max}p_1 M/\sqrt{I_{FZP}})^2$ for $\sqrt{I_{FZP}}/(Ma_{0max} + Ma_{0min}) \leq p_1 \leq 0.5$. By visualizing $SR_{min}$ and $SR_{max}$, Figure 1c illustrates the correlation between the Strehl ratio and the information entropy $S$ by using an intermediate parameter $p_1$. With increasing $S$, both $SR_{min}$ and $SR_{max}$ decrease, implying large aberration and poor imaging quality. However, its increased range ($\Delta SR = SR_{max} - SR_{min}$) indicates high uncertainty of $SR$, which has echoes of high disorder for a large $S$. Therefore, a small $S$ with low disorder is preferred in imaging, which needs a large SR.

Similarly, its focal spot is also controlled by the entropy $S$ or $p_1$. Although it is difficult to derive the spot size analytically, we obtain its upper ($r_{max}$) and lower ($r_{min}$) boundaries numerically (see Supplementary Section 1), as illustrated in Fig. 1c. For the larger entropy $S$, the spot size is valued in the wider range, which offers more opportunities to realize super-focusing. Hence, a large entropy $S$ with high disorder is required in super-focusing. These results reveal that imaging and super-focusing have completely opposite requirement for the entropy $S$ of an objective chip, which doubly confirms that the entropy $S_0 = 0.5$ leads to good balance.

**Design of objective chip**

Since the FZP contained in the objective chip is given with analytically described structures, we only need to optimize its deviation part $U_\Delta = e^{i\Delta\varphi}$, which refers to a few-ring phase mask (see Fig. 1b). Considering that the ideal equilibrium point $S_0 = 0.5$ has a small deviation probability, only 5 rings are used to avoid introducing high disorder. In our design, all radii $\rho_n$ ($n=1, 2, ..., 5$) of these 5 rings are chosen from the radii of zones in the FZP, which means that no new finer structure is created when combining the FZP and this 5-ring phase mask. Such a design strategy has twofold significance. First, it greatly enhances the speed of optimization. Because all the structural details of the objective chip are given by the radii $r_m$ of the FZP, we can calculate the electric field of each zone ahead of optimization and then store it as a database for quick reading during the design, thus avoiding repeated calculation. For example, to design our objective chip with 6387 zones, it takes only 6 mins to run 500 iterations by using the particle swarm algorithm in a personal computer, enhancing the speed by a factor of ~$3.8 \times 10^5$ (see the calculation details in the Methods). Second, it allows large-scale and low-cost fabrication of our objective chip because its smallest feature size is theoretically fixed to $\Delta r_{min} = \lim_{m \to \infty}(r_m - r_{m-1}) = \lambda/2$.

The 5-ring phase mask in our objective chip is designed with $\rho_1 = r_{283}$, $\rho_2 = r_{850}$, $\rho_3 = r_{1046}$, $\rho_4 = r_{1258}$ and $\rho_5 = r_{6391}$, which yields $p_1$=(567+212)/6391=0.122 and entropy $S$=0.535. Compared with other reported planar lenses[9, 13, 14, 22-25], the achieved entropy in our objective chip is the closest to the equilibrium point $S_0$=0.5 (see Fig. 1d), implying its advantages in balancing optical imaging and super-focusing. It exhibits the fundamental difference of our objective chip from other lenses with their entropies approaching 1. Furthermore, the relative $SR$=0.45 of our objective chip is also the highest among those lenses, which theoretically predicts the best imaging capability. The optimized focal spot has a lateral FWHM (*i.e.*, full width at half maximum) of 180 nm (0.44$\lambda$, below the Rayleigh criterion of $r_{RC}$=0.51$\lambda$/$NA$=0.57$\lambda$), which is slightly larger than the superoscillation criterion $r_{SOC}$=0.358$\lambda$/$NA$=0.398$\lambda$ (in terms of FWHM)[26] to avoid strong sidebands in a superoscillatory spot. The simulated electric fields near the focal plane are provided with experimental results for a better comparison, as shown later.

**Fabrication of objective chip**

The possibility of employing semiconductor processes to fabricate planar flat optics will ultimately allow mass production of flat optics at a low cost and push this technique for wide market adoption. As discussed above, our design strategy makes low-cost and fast DUV lithography feasible for fabricating a 4-mm-diameter objective chip. Under the conditions of $\lambda$=405 nm, $f$=1 mm and $NA$=0.9 in this work, $\Delta r_{min}$ is only 225 nm, which is within the capability of commercial DUV lithography (with a critical dimension of 200 nm). This key feature will greatly facilitate the future manufacturing of our objective chips by reducing the cost compared with other flat lenses with critical dimensions smaller than 200 nm that would require much more costly 12-inch immersion lithography[27]. We note that E-beam lithography is able to write the pattern at high resolution; however, it is not the technique for large-scale fabrication due to its speed limit and the inevitable stitching error (with a small writing square of several tens of micrometers). The fabrication details of our objective chip are provided in the Methods.

The inset of Fig. 1a shows the microscopic image of our fabricated objective chip, where different colors arise from optical scattering of daylight. To reveal the fine details, we show scanning-electron-microscopy (SEM) images of the objective chip in Fig. 2a, where both simulated and experimental widths from the center to the outermost boundary are also provided with a maximum deviation of <75 nm. The possible reason comes from insufficient exposure time of photoresist under DUV radiation, which can be solved by increasing the exposure time. Using a profilometer, we characterize the groove depth of ~530 nm (see the insert in Fig. 2a) that yields a phase modulation of ~1.23π, which is larger than the ideal value of π due to overetching issues. Nevertheless, we emphasize that such a deviation of 0.23π in phase modulation leads to a theoretical decrement of only ~2.6% in the focusing efficiency of the objective chip, see the simulation details in the Methods.

**Subdiffraction-limit focusing by objective chip**

To verify the focusing capability of the objective chip, we first measure the diffraction field near the focus of the objective chip under the illumination of collimated circularly polarized light by using a 0.95-*NA* objective lens, see the experimental details in Supplementary Section 3. Figure 2b shows the *x-z* and *y-z* cross-section of the measured light intensity, which reveals a record-long focal length of 1 mm (compared with the previous planar diffractive lenses[9, 14, 15, 22, 28, 29]). The axial depth of focus (DOF) is extended from the simulated ~500 nm to the experimental ~5500 nm, which is caused by the fabrication error of the groove width in the objective chip (see Fig. 2a). Such a long DOF offers good tolerance to sample alignment in an SCM[13]. Fortunately, the focusing spot size of the objective chip is less influenced by the weak phase perturbation from the imperfect zones, as confirmed by the good agreement between the simulated and experimental line-scanning intensity profiles (Fig. 2c). To quantitatively evaluate the focusing effect, we present the experimental FWHM of the focal spot size along the propagation of light in Fig. 2d, indicating the varying FWHM from 170 nm to 210 nm (tightly close to the simulated 180 nm). Compared with the Rayleigh criterion of 0.51*λ*/*NA*, these achieved focal spots confirm the subdiffraction-limit focusing capability of the proposed objective chip. Due to its supercritical feature with lateral FWHMs above the superoscillation criterion[26], the focusing spots have no strong sideband, as observed in Supplementary Video 1, which dynamically records the focusing process near the focal plane.

The focusing efficiency, defined as the ratio of the focused power (experimentally filtered by a 150 μm-diameter pinhole at the focal plane) to the total power incident on the objective chip, is measured to be 12.3% (see the measurement details in Supplementary Section 4), which is lower than its theoretical efficiency of 18.7% (see its calculation in the Methods), as shown in Fig. 2e. This discrepancy in efficiency is attributed to incomplete constructive interference of light diffracted from two neighboring zones because of the insufficient etching widths. Although our achieved efficiency is not as high as those of traditional objectives and metalenses, it still exhibits significant enhancement in comparison with those of amplitude-type planar diffractive lenses[9, 14, 15, 22, 29].

**Direct wide-field imaging by objective chip**

To investigate its wide-field imaging, a knife-edge object (see its microscopic image at the left-bottom of Fig. 2f) is located at a distance of $z=1.2f$ from the objective chip[30]. After illuminating the knife-edge object, the transmitted light is collected by our objective chip. According to the imaging formula of a lens, we can roughly estimate its imaging distance of $6f$ at the other side of the objective chip, thereby exhibiting an imaging magnification of 5×. The recorded image of the knife-edge object is shown at the bottom-right panel of Fig. 2f, which reveals a well-defined boundary at the edge. A dynamic video that records its imaging process by tuning the axial position of such a knife-edge object is provided in Supplementary Video 2, which is captured in a homemade measurement setup (see Supplementary Fig.8 and Supplementary Section 5).

The modulation transfer function (MTF) of this objective chip is characterized by using the line-scanning intensity across the edge in the image. To decrease the experimental error, we employ a mean of the line-scanning intensity (with its spatial dimension scaled down by a factor of its magnification $M=5$) at the red-rectangle region in the image of Fig. 2f to recover the line spread function (LSF) of this imaging configuration. After fitting the line-scanning intensity with an error function (taken as a convolution between a Gaussian function and a jump function), we carry out the deconvolution of the fitted error function, yielding the retrieved LSF (Fig. 2g). Using a Fourier transformation of the retrieved LSF, we finally obtain its MTF (Fig. 2h), which indicates a cut-off frequency of 4000 lp/mm. This implies an imaging resolution of 250 nm, which corresponds to an effective $NA$ of 0.83 (evaluated by using $0.51\lambda/NA_{eff}=250$ nm) for imaging. Compared with the previous metalens with a cut-off spatial frequency of 2000 lp/mm [4], our objective chip achieves a twofold enhancement in resolving power when operating in the mode of direct wide-field imaging. These experimental results have confirmed that our objective chip has sufficient imaging ability to collect high spatial frequencies from fine details of objects, which is superior to all previous superoscillation[9, 10, 12, 22, 28, 29] and supercritical lenses[13-15].

**Objective-chip-based reflective SCM**

A high-resolution reflective SCM (Fig. 3a, see its working principle in the Methods) has been built successfully due to both features (*i.e.,* subdiffraction-limit focusing and direct wide-field imaging) of our objective chip. First, the enhanced focusing efficiency allows more light to illuminate the object in a reflective mode, which underpins subsequent collection and detection of reflected light. Figure 3b shows the experimental signals detected by the photomultiplier tube (PMT) and charge-coupled device (CCD) when the nano-objects are moved longitudinally near the focal plane of $z=1$ mm. It reveals that the PMT signal reaches its maximum for the in-focus (*i.e.*, $\Delta z=0$) nano-objects and decreases gradually with the increment of the out-of-focus distance $|\Delta z|$, which is doubly checked by the CCD images (see the inserts in Fig. 3b and Supplementary Video 3). The FWHM of the PMT measured intensity profile is ~5 μm, which agrees with the experimental DOF of 5.5 μm (see Figs. 2b and 2d). This result confirms that our objective chip can efficiently focus the incident beam and collect the reflected light. Second, a powerful imaging ability with an effective *NA* of 0.83 is required to enhance the practical imaging resolution of an SCM. Theoretically, we have already shown that the resolution of an SCM is less influenced by the *NA* of the collection objective[7], which, however, is valid only for infinitesimal point objects. For real objects with finite sizes ranging from tens to hundreds of nanometers, the *NA* of the collection objective should be larger than 0.7 for a better resolution in an SCM, see the simulated proofs in Supplementary Section 6. Due to these two features of our objective lens mentioned above, the subdiffraction-limit focusing and the high-resolution imaging enable a reflective SCM. More experimental details about the scanning imaging are provided in the Methods.

To test its resolution, we provide the imaging results of 50 nm-width and $2\mu m$-length double slits with center-to-center (CTC) distances ranging from 190 nm to 270 nm. As shown in Fig. 3c, these slits are etched on a 140 nm-thick chromium film on a quartz substrate. Using a 0.9 NA objective for a fair comparison, the coherent bright-field microscope cannot resolve these double slits (Fig.3d) while conventional reflective SCM can only resolve the double slits with CTC distances larger than 240 nm (Fig.3e). In contrast, our objective-chip-based reflective CSM has an enhanced resolution so that double slits with a CTC distance of 200 nm can be distinguished with a valley of intensity in the image (Fig.3f), where the distortion is caused by mechanical variation of the sample. The qualitative comparisons among their line-scanning intensity profiles (Fig. 3g) doubly verify an imaging resolution of 200 nm achieved by our SCM. In addition, all these experimental results regarding scanning images are confirmed by our simulations (see Supplementary Section 7 and Supplementary Fig. 10) with the theory of SCM[7, 31, 32].

Complex nano-objects can also be imaged with a high resolution by using our SCM. Figure 4a shows the SEM image of a dolphin (composed of 50 nm-width curves) with a total size of 8 μm×8 μm. Due to their limited resolutions, both coherent BF microscopy and traditional SCM can map only rough contours of the dolphin but lose fine details, such as the eye and tail (Figs. 4b and 4c). In comparison, our SCM can clearly resolve all these fine details (Fig. 4d) with a narrower line width (Fig. 4e). Furthermore, two lines with a CTC distance of 225 nm (see the dashed-red lines at the lower rows in Figs. 4b-4d) in the tail can be distinguished only by using our SCM. The low contrast of intensity in the image for our SCM comes from the relatively lower focusing efficiency of the objective chip in comparison with the traditional objective. However, it has little influence on the resolution and clarity of the image, as observed in Fig. 4d.

**Discussion**

Among all the planar-lens-based SCMs, our current SCM has the advantages of eliminating bulky objectives, a millimeter-level working distance, reflection-mode operation, working for both transparent and non-transparent substrates, and a competitive resolution of 0.49λ, as shown in Supplementary Table 1. For commercial objectives, pursuing a high *NA* and long working distance simultaneously leads to an increment in the diameters of optical elements and the accompanying optical aberrations that need large-scale nonspherical surfaces for correction[3], thereby yielding extremely high costs in both the fabrication and design of elements. This issue does not exist in our objective chip, where the smallest feature of *λ*/2 will not change with increasing *NA* and focal length, so the same fabrication tools and design methods reported in this work can also be used to develop more advanced objective chips with even larger *NA*s and longer focal lengths.

Our objective chip has a fabrication cost of ~$42 dollars (estimated by the total price of 300 chips in an 8-inch quartz wafer, see Supplementary Fig. 11), which is ~100 times cheaper than the price of commercial objectives. This objective chip has a volume of 4 mm×4 mm×0.5 mm, which indicates a shrinking factor of 4300 (~3.6 orders of magnitude) compared with traditional objectives (ZEISS, EC Epiplan-Neufluar 100×, *NA* 0.9, M27).

Developing this reflective configuration makes a truly important step to push the technology of planar-lens-based SCMs towards practical applications, because many samples have opaque substrates that are incompatible with all the previous planar-lens-based SCMs. Note that the objects used in this work are made in a high-reflectivity metal film, which helps to enhance the imaging contrast. If the difference between the optical reflection of the object and its surrounding background is not obvious, one should increase the optical efficiency of the objective chip and the sensitivity of the optical detector. The focusing efficiency of the objective chip can be enhanced further if multilevel phase elements[33] or high-efficiency dielectric metasurfaces[4, 5, 34-43] are used. The detector can also be updated to the single-photon level for a better recording of collected photons by our objective chip, enabling the characterization of less-reflective biological tissues and cells even in a living body.

In summary, we have proposed information entropy to evaluate the disorder of an optimized planar lens. The suggested equilibrium point $S_0$=0.5 is used to guide the quick design of a 1-mm focal-length, high-*NA* and low-cost objective chip with efficiency-enhanced subdiffraction-limit focusing and direct wide-field imaging. These advantages open the way to demonstrate compact and high-resolution reflective SCM with planar lenses, which will greatly benefit from optical to biomedical imaging.

## Methods

**Design and optimization details.** To avoid the creation of additional finer structures when combining the FZP and the few-ring mask, each radius $\rho_n$ (*n*=0, 1, 2, … , *N*) in the *N*-ring mask is valued within the radii (*i.e.*, $r_m$) of belts in the FZP. To highlight these selected radii in the zone plates, we label them $r_{M_n} = \rho_n$, which means that the $n^{\text{th}}$ ring of the few-ring mask has the same radius as the $M_n^{\text{th}}$ belt of the zone plate. Although it decreases the degree of freedom to design the few-ring mask, significant benefits are achieved in simplifying the optimization and fabricating the sample.

Benefiting from this design strategy, we can express the electric field of light focused by the objective chip as

$$E_{chip}(u,v,z) = \sum_{n=0}^{N-1}(-1)^n \left[\sum_{m=M_n}^{m=M_{n+1}}(-1)^m A_m\right] = \sum_{n=0}^{N-1}\sum_{m=M_n}^{m=M_{n+1}}(-1)^{m+n} A_m$$

$$= \sum_{n=0}^{N-1}\sum_{m=M_n}^{m=M_{n+1}}(-1)^{m+n} \int_{r_m}^{r_{m+1}}\int_0^{2\pi} E_0(r,\varphi)\frac{exp(ikR)}{R^2}\left(ik-\frac{1}{R}\right)zrdrd\varphi, \quad (3)$$

where $A_m$ is the electric field of light diffracting from the $m^{\text{th}}$ belt in the FZP, the wavenumber $k$=2π/λ,

$R^2=(x-v)^2+(y-u)^2+z^2$, $r^2=x^2+y^2$, $tan\varphi=y/x$, $x$ and $y$ are the spatial coordinates at the plane of the objective chip, $x$, $y$ and $z$ stand for the spatial position of interest. Considering the nonparaxial propagation of light in such a high-$NA$ objective chip, we calculate $A_m$ by using the rigorous Rayleigh-Sommerfeld diffraction integral without any approximation[7, 44]. Although thousands of belts are included in this objective chip, only $N$-1 variables (*i.e.*, $M_1$, $M_2$, …, $M_{N-1}$ because $M_0$=0 and $M_N$=6391) are unknown in Eq. (3) because all $r_m$ are given. Since the phase jump of $\pi$ occurs at the $M_n^{th}$ belt ($n$=1, 2, …, $N$-1), both the $M_n^{th}$ and ($M_n$+1)$^{th}$ belts are combined into one, leading to the belt number of $M$-($N$-1) in the final objective chip.

Since no new belt appears in this strategy, the electric field $A_m$ can be calculated ahead of optimization and then stored in a database, thus enhancing the design speed. The particle swarm algorithm[45] is used to optimize the $N$-1 parameters, see the details in Supplementary Section 2. In our design, $N$=5 is employed with 4 unknown parameters, which can be determined with the values of $M_1$=283, $M_2$=850, $M_3$=1046 and $M_4$=1258 by carrying out 500 iterations in ~6 minutes in a personal computer (Intel Core i5-7500 CPU 3.40 GHz, 32G RAM). In each iteration, the optimization algorithm contains 20 populations, each of which stands for one design of objective chip. If our design strategy with the pre-calculated database is not used, we can roughly estimate its time cost of 3.8×10$^4$ (=3.8×20×500) hours to finish the design by running 500 iterations, because it will take ~3.8 hours to calculate the focal field of a single objective chip by numerically integrating all the zones with Rayleigh-Sommerfeld diffraction under the same computation environment. Thus, our design strategy accelerates the optimization by a factor of 3.8×10$^5$. In our designed objective chip, the phase-reversed zones contain two parts: 1) from $m$=284 to $m$=850 and 2) from $m$=1047 to $m$=1258, resulting in $p_1$=(567+213)/6391=0.122 and $p_2$=1-$p_1$=0.878. According to the definition of information entropy, we have $S$=0.535, which is tightly close to the equilibrium point $S_0$=0.5.

**Fabrication details.** The designed objective chip is fabricated through deep-ultraviolet (DUV, Nikon S204) exposure process. The quartz substrate is first deposited with 200 nm-thick aluminum using a physical vapour deposition (PVD) system (AMAT Endura). Then, a 300 nm-thick positive photoresist (UV135) is coated and baked. Subsequently, the photoresist is patterned using the DUV lithography. After development, the aluminum film without photoresist is etched sufficiently by an inductively coupled plasma (ICP) etching system (LAM 9600). Thus, the patterns of the objective chip are transferred into the aluminum film after removing the residual photoresist. Next, using the aluminum film as masking layer, the quartz substrate is etched for a designed thickness by an inductively coupled plasma-reactive ion etching (ICP-RIE) system (Oxford, Plasma Pro System100 ICP380). Finally, the aluminum hard mask is removed by Tetramethylammonium Hydroxide (TMAH, 2.5%) solvent, yielding the expected phase-type objective chip.

**Theoretical efficiency of the objective chip.** To obtain the theoretical efficiency of the objective chip, we first calculate the electric field $A_m$ of the $m^{th}$ belt in the zone plate at the focal plane (ignoring the influence from the width error of etched belts). For the sake of convenient simulation, we calculate the one-dimensional field along the radial direction within the range of λ (starting from $r$=0), where the focused light is concentrated. To evaluate the experimental error, we update Eq. (3) by considering an actual phase difference of $\Delta \varphi$ as

$$E_{chip}(r) = \sum_{n=0}^{n=4} \sum_{m=M_n}^{m=M_{n+1}} e^{i \cdot \Delta \varphi \cdot \frac{1+(-1)^{m+n}}{2}} A_m, \qquad (4)$$

where $M_0$=0, $M_1$=283, $M_2$=850, $M_3$=1046, $M_4$=1258, and $M_5$=6391, $\Delta \varphi = \frac{2\pi}{\lambda}(n-1)d_{etch}$ is the phase difference between etched and unetched zones in the objective chip at a wavelength of $\lambda$=405 nm, the refraction index of the quartz substrate is $n$=1.47 and the experimental etching depth $d_{etch}$=530 nm (referring to $\Delta \varphi = 1.23\pi$). To further evaluate the energy flux in the circular area with a radius of λ at the focal plane of the objective chip, we employ the expression:

$$W_{chip} = \int_0^\lambda \int_0^{2\pi} I_{chip}(r) r dr d\varphi, \qquad (5)$$

where the intensity $I_{chip}(r) = |E_{chip}(r)|^2$. Similarly, we can acquire the energy flux of the standard binary-phase zone plate in the same area with $W_{FZP} = \int_0^\lambda \int_0^{2\pi} I_{FZP}(r) r dr d\varphi$, where $I_{FZP} = |E_{FZP}(r)|^2 = |\sum_{m=0}^{m=6391}(-1)^m A_m|^2$. Finally, the theoretical focusing efficiency of the objective chip can be evaluated as

$$\eta_{chip} = \eta_{FZP} \cdot \frac{W_{chip}}{W_{FZP}}, \qquad (6)$$

where $\eta_{FZP} = sinc^2(1/L)$ denotes the optical efficiency of multilevel phase elements, and $L$ is the number of phase levels. For the binary-phase FZP, we have $\eta_{FZP} = 40.5\%$. Note that, Eq. (6) depends on the geometric parameters $M_n$ and $\Delta\varphi$, which allows us to conveniently investigate optical efficiency of the objective chip. For example, once the geometric parameters are fixed in this work, we can simulate the theoretical focusing efficiency for different $\Delta\varphi$ values (see Fig. 2e), which reveals a peak efficiency of 21.3% at $\Delta\varphi$=π. For the experimental $\Delta\varphi$=1.23π, its corresponding theoretical efficiency is 18.7% with a deviation of 2.6% from the peak efficiency.

**Work principle of objective-chip-based reflective SCM.** A schematic diagram of the objective-chip-based reflective SCM is given in Fig. 3a. The confocal configuration consists of our objective chip and two tube lenses ($TL_1$ and $TL_2$), where their focal planes are conjugated with that of the objective chip. The objective chip is illuminated by a collimated light beam with a wavelength $\lambda=405$ nm.

To increase the signal-to-noise ratio of the entire system, we suppress the light reflected from the back-surface (*i.e.*, the bare-quartz side without any structure) of the objective chip by utilizing two orthogonal linear polarizers ($LP_1$ and $LP_2$) and a quarter-waveplate (QWP) thin film, as sketched in Fig. 3a. Because both $LP_1$ and $LP_2$ have orthogonal transmission directions, the reflected light from the back-surface of the objective chip is blocked efficiently, leaving a very weak background with a four-lobe pattern (see the insert in Fig. 3b). It is important to note that such a four-lobe pattern has a dark center, where the signal light reflected by the objects is located, thus leading to spatial separation between the noise and signal. The 60 μm-thick QWP thin film (with an angle of 45 degrees between its fast axis and the transmission directions of both LPs) is adhered to the structure side of the objective chip to obtain circular polarization (CP), which enables us to realize a circularly symmetric focal spot for isotropic scanning of the image. Moreover, the reflected CP signal light from nano-objects passes through the QWP thin film again and is converted into linear polarization with its direction aligned to the transmission direction of $LP_2$. Therefore, the second linear polarizer ($LP_2$) can block the noise light reflected from the back surface of the objective chip and transmit the signal light efficiently, thus increasing the signal-to-noise ratio of the PMT signals.

After tuning the signal, we coarsely move the objective chip mounted on the electric stage toward the nano-objects. When the nano-objects are close to the focal plane, the PMT signals behave like that shown in Fig. 3b, having a peak when the nano-objects are in focus. At the same time, the recorded pattern at the CCD becomes the smallest. Due to the conjugation relationship between the objects and the PMT, we can adjust the PMT to collect an optical signal at the focal plane of $TL_2$ filtered by a 10 μm-diameter pinhole. Then, we finely move the nano-objects mounted on the 3D piezo stage to the focal plane by observing the signal collected by the PMT. When the PMT signal reaches its maximum (see Fig. 3b), we assume that the nano-objects are at the focal plane of the objective chip. Finally, the nano-objects can be scanned at the focal plane, and the signal collected by the PMT can be recorded simultaneously to complete the scanning image.

**Experimental details about the scanning imaging.** The 3D piezo stage (PI-545.3R8S) and the controller (PI-E727) are integrated into a single device with a scanning resolution of ~1 nm. The PMT has the module of the WiTec 3000R series. We utilize the LabVIEW language to control the movement of the 3D piezo stage and read the signal collected by the PMT with a DAQ card (NI USB-6000, 12-bit, sampling rate: 10 Ks/s) simultaneously. A $10\,\mu m$-diameter fiber (Thorlab M64L01 $10\mu m$ 0.1 NA) is employed here as the pinhole.

In the experiment testing the imaging resolution, the scanning range of double slits is 3 μm×1 μm with 100×100 pixels, which takes ~15 minutes to finish one image. The scanning speed can be updated further by using a high-speed stage and digital-analog converter. The 2.4 μm×0.6 μm range of scanning images in Figs. 3(c)-(f) is employed to fully cover the objects. In the experiment for imaging complex nano-objects, the scanning range is 9 μm×9 μm with 150×150 sampling points, which takes ~34 minutes. Only the 8 μm×8 μm range is shown in Figs. 4(b)-(d) to highlight more details of the images.

**Acknowledgement**

K.H. thanks the CAS Project for Young Scientists in Basic Research (Grant No.YSBR-049), the National Natural Science Foundation of China (Grant No. 12134013), the National Key Research and Development Program of China (No. 2022YFB3607300), the CAS Pioneer Hundred Talents Program, and support from the University of Science and Technology of China's Centre for Micro and Nanoscale Research and Fabrication. J.T. thanks the A*STAR AME IRG program (Grant No. A2083c0058).

**Author contributions**

K. H., J. T. and C. Q. conceived the idea. J. H. and K. H. performed the simulations. D. Z. and H. L. prepared and fabricated optical samples. J. H. and D. Z. built up the experimental setup and performed the characterization. J. H., K. H., C. Q. and J. T. wrote the manuscript. K. H., J. T., and C. Q. supervised the overall project. All authors discussed the results, carried out the data analysis and commented on the manuscript.

**Competing financial interests**

The authors declare no competing financial interests.

**References**

[1]. Croft, W. J. *Under the microscope: a brief history of microscopy*.(World Scientific, **2006**).

[2]. Keller, H. E. *Handbook of biological confocal microscopy*.(Springer, **2006**).

[3]. Born, M., & Wolf, E. *Principles of optics: electromagnetic theory of propagation, interference and diffraction of light*.(Cambridge Univ. Press, **1999**).

[4]. Khorasaninejad, M., Chen, W. T., Devlin, R. C., Oh, J., Zhu, A. Y., & Capasso, F., Metalenses at visible wavelengths: Diffraction-limited focusing and subwavelength resolution imaging, *Science* **352**, 1190(2016).

[5]. Huang, K., Deng, J., Leong, H. S., Yap, S. L. K., Yang, R. B., Teng, J., & Liu, H., Ultraviolet Metasurfaces of ≈80% Efficiency with Antiferromagnetic Resonances for Optical Vectorial Anti-Counterfeiting, *Laser & Photonics Rev.* **13**, 1800289(2019).

[6]. Lalanne, P., & Chavel, P., Metalenses at visible wavelengths: past, present, perspectives, *Laser Photonics Rev.* **11**, 1600295(2017).

[7]. Huang, K., Qin, F., Hong Liu, Ye, H., Qiu, C. W., Hong, M., Luk'yanchuk, B., & Teng, J., Planar Diffractive Lenses: Fundamentals, Functionalities, and Applications, *Adv. Mater.* **30**, 1704556(2018).

[8]. Chen, W. T., Zhu, A. Y., Khorasaninejad, M., Shi, Z., Sanjeev, V., & Capasso, F., Immersion Meta-Lenses at Visible Wavelengths for Nanoscale Imaging, *Nano Lett.* **17**, 3188(2017).

[9]. Rogers, E. T., Lindberg, J., Roy, T., Savo, S., Chad, J. E., Dennis, M. R., & Zheludev, N. I., A super-oscillatory lens optical microscope for subwavelength imaging, *Nat. Mater.* **11**, 432(2012).

[10]. Huang, K., Liu, H., Garcia-Vidal, F. J., Hong, M., Luk'yanchuk, B., Teng, J., & Qiu, C.-W., Ultrahigh-capacity non-periodic photon sieves operating in visible light, *Nat. Commun.* **6**, 7059(2015).

[11]. Huang, F. M., & Zheludev, N. I., Super-Resolution without Evanescent Waves, *Nano Lett.* **9**, 1249(2009).

[12]. Tang, D., Wang, C., Zhao, Z., Wang, Y., Pu, M., Li, X., Gao, P., & Luo, X., Ultrabroadband superoscillatory lens composed by plasmonic metasurfaces for subdiffraction light focusing, *Laser Photon. Rev.* **9**, 713(2015).

[13]. Qin, F., Huang, K., Wu, J., Teng, J., Qiu, C.-W., & Hong, M., A Supercritical Lens Optical Label-Free Microscopy: Sub-Diffraction Resolution and Ultra-Long Working Distance, *Adv. Mater.* **29**, 1602721(2017).

[14]. Wang, Z., Yuan, G., Yang, M., Chai, J., Steve Wu, Q. Y., Wang, T., Sebek, M., Wang, D., Wang, L., Wang, S., Chi, D., Adamo, G., Soci, C., Sun, H., Huang, K., & Teng, J., Exciton-Enabled Meta-Optics in Two-Dimensional Transition Metal Dichalcogenides, *Nano Lett.* **20**, 7964(2020).

[15]. Qin, F., Liu, B., Zhu, L., Lei, J., Fang, W., Hu, D., Zhu, Y., Ma, W., Wang, B., Shi, T., Cao, Y., Guan, B.-o., Qiu, C.-w., Lu, Y., & Li, X., π-phase modulated monolayer supercritical lens, *Nat. Commun.* **12**, 32(2021).

[16]. Rogers, E. T. F., & Zheludev, N. I., Optical super-oscillations: sub-wavelength light focusing and super-resolution imaging, *J. Opt.* **15**, 094008(2013).

[17]. Chen, G., Wen, Z.-Q., & Qiu, C.-W., Superoscillation: from physics to optical applications, *Light: Sci. Appl.* **8**, 56(2019).

[18]. Berry, M., Zheludev, N., Aharonov, Y., Colombo, F., Sabadini, I., Struppa, D. C., Tollaksen, J.,

Rogers, E. T. F., Qin, F., Hong, M., Luo, X., Remez, R., Arie, A., Götte, J. B., Dennis, M. R., Wong, A. M. H., Eleftheriades, G. V., Eliezer, Y., Bahabad, A., Chen, G., Wen, Z., Liang, G., Hao, C., Qiu, C. W., Kempf, A., Katzav, E., & Schwartz, M., Roadmap on superoscillations, *J. Opt.* **21**, 053002(2019).

[19]. Levine, A., & Markowitz, O., Introduction to reflectance confocal microscopy and its use in clinical practice, *JAAD Case Rep.* **4**, 1014(2018).

[20]. Ayres, R. U. *Information, entropy, and progress: a new evolutionary paradigm*.(Springer Science & Business Media, **1997**).

[21]. Huang, K. *Introduction to statistical physics*.(Chapman and Hall/CRC, **2009**).

[22]. Yuan, G. H., Rogers, E. T., & Zheludev, N. I., Achromatic super-oscillatory lenses with sub-wavelength focusing, *Light: Sci. Appl.* **6**, e17036(2017).

[23]. Yu, A.-p., Chen, G., Zhang, Z.-h., Wen, Z.-q., Dai, L.-r., Zhang, K., Jiang, S.-l., Wu, Z.-x., Li, Y.-y., & Wang, C.-t., Creation of sub-diffraction longitudinally polarized spot by focusing radially polarized light with binary phase lens, *Sci. Rep.* **6**, 38859(2016).

[24]. Yuan, G., Rogers, E. T., Roy, T., Adamo, G., Shen, Z., & Zheludev, N. I., Planar super-oscillatory lens for sub-diffraction optical needles at violet wavelengths, *Sci. Rep.* **4**, 6333(2014).

[25]. Qin, F., Liu, B., Zhu, L., Lei, J., Fang, W., Hu, D., Zhu, Y., Ma, W., Wang, B., & Shi, T., π-phase modulated monolayer supercritical lens, *Nat. Commun.* **12**, 32(2021).

[26]. Huang, K., Ye, H., Teng, J., Yeo, S. P., Luk'yanchuk, B., & Qiu, C. W., Optimization-free superoscillatory lens using phase and amplitude masks, *Laser Photon. Rev.* **8**, 152(2014).

[27]. Park, J.-S., Zhang, S., She, A., Chen, W. T., Lin, P., Yousef, K. M. A., Cheng, J.-X., & Capasso, F., All-Glass, Large Metalens at Visible Wavelength Using Deep-Ultraviolet Projection Lithography, *Nano Lett.* **19**, 8673(2019).

[28]. Yuan, G. H., Lin, Y.-H., Tsai, D. P., & Zheludev, N. I., Superoscillatory quartz lens with effective numerical aperture greater than one, *Appl. Phys. Lett.* **117**, 021106(2020).

[29]. Yuan, G., Rogers, K. S., Rogers, E. T. F., & Zheludev, N. I., Far-Field Superoscillatory Metamaterial Superlens, *Phys. Rev. Appl.* **11**, 064016(2019).

[30]. Tzannes, A. P., & Mooney, J. M., Measurement of the modulation transfer function of infrared cameras, *Opt. Eng.* **34**, 1808(1995).

[31]. Wilson, T., & Sheppard, C. *Theory and practice of scanning optical microscopy*.(Academic Press London, **1984**).

[32]. Sheppard, C. J. R., & Choudhury, A., Image-Formation in Scanning Microscope, *Opt. Acta.* **24**, 1051(1977).

[33]. Banerji, S., Meem, M., Majumder, A., Vasquez, F. G., Sensale-Rodriguez, B., & Menon, R., Imaging with flat optics: metalenses or diffractive lenses?, *Optica* **6**, 805(2019).

[34]. Zhang, C., Divitt, S., Fan, Q., Zhu, W., Agrawal, A., Lu, Y., Xu, T., & Lezec, H. J., Low-loss metasurface optics down to the deep ultraviolet region, *Light: Sci. Appl.* **9**, 1(2020).

[35]. Fan, Z.-B., Shao, Z.-K., Xie, M.-Y., Pang, X.-N., Ruan, W.-S., Zhao, F.-L., Chen, Y.-J., Yu, S.-Y., & Dong, J.-W., Silicon Nitride Metalenses for Close-to-One Numerical Aperture and Wide-Angle Visible Imaging, *Phys. Rev. Appl.* **10**, 014005(2018).

[36]. Arbabi, A., Arbabi, E., Kamali, S. M., Horie, Y., Han, S., & Faraon, A., Miniature optical planar camera based on a wide-angle metasurface doublet corrected for monochromatic aberrations, *Nat. Commun.* **7**, 13682(2016).

[37]. Arbabi, A., Horie, Y., Ball, A. J., Bagheri, M., & Faraon, A., Subwavelength-thick lenses with high

numerical apertures and large efficiency based on high-contrast transmitarrays, *Nat. Commun.* **6**, 7069(2015).

[38]. Arbabi, A., Horie, Y., Bagheri, M., & Faraon, A., Dielectric metasurfaces for complete control of phase and polarization with subwavelength spatial resolution and high transmission, *Nat. Nanotechnol.* **10**, 937(2015).

[39]. Wang, S., Wu, P. C., Su, V.-C., Lai, Y.-C., Chen, M.-K., Kuo, H. Y., Chen, B. H., Chen, Y. H., Huang, T.-T., Wang, J.-H., Lin, R.-M., Kuan, C.-H., Li, T., Wang, Z., Zhu, S., & Tsai, D. P., A broadband achromatic metalens in the visible, *Nat. Nanotechnol.* **13**, 227(2018).

[40]. Wang, S., Wu, P. C., Su, V.-C., Lai, Y.-C., Hung Chu, C., Chen, J.-W., Lu, S.-H., Chen, J., Xu, B., Kuan, C.-H., Li, T., Zhu, S., & Tsai, D. P., Broadband achromatic optical metasurface devices, *Nat. Commun.* **8**, 187(2017).

[41]. Chen, W. T., Zhu, A. Y., Sanjeev, V., Khorasaninejad, M., Shi, Z., Lee, E., & Capasso, F., A broadband achromatic metalens for focusing and imaging in the visible, *Nat. Nanotechnol.* **13**, 220(2018).

[42]. Fu, W., Zhao, D., Li, Z., Liu, S., Tian, C., & Huang, K., Ultracompact meta-imagers for arbitrary all-optical convolution, *Light: Sci. Appl.* **11**, 62(2022).

[43]. Fu, W., Zhao, D., Li, Z., & Huang, K., Optical Magnetism-Induced Dual Anisotropy in Dielectric Nanoantennas, *Adv. Opt. Mater.* **10**, 2200303(2022).

[44]. Huang, K., Liu, H., Si, G., Wang, Q., Lin, J., & Teng, J., Photon-nanosieve for ultrabroadband and large-angle-of-view holograms, *Laser Photon. Rev.* **11**, 1700025(2017).

[45]. Jin, N., & Rahmat-Samii, Y., Advances in particle swarm optimization for antenna designs: Real-number, binary, single-objective and multiobjective implementations, *IEEE Trans. Antennas Propag.* **55**, 556(2007).

**Figures and Captions**

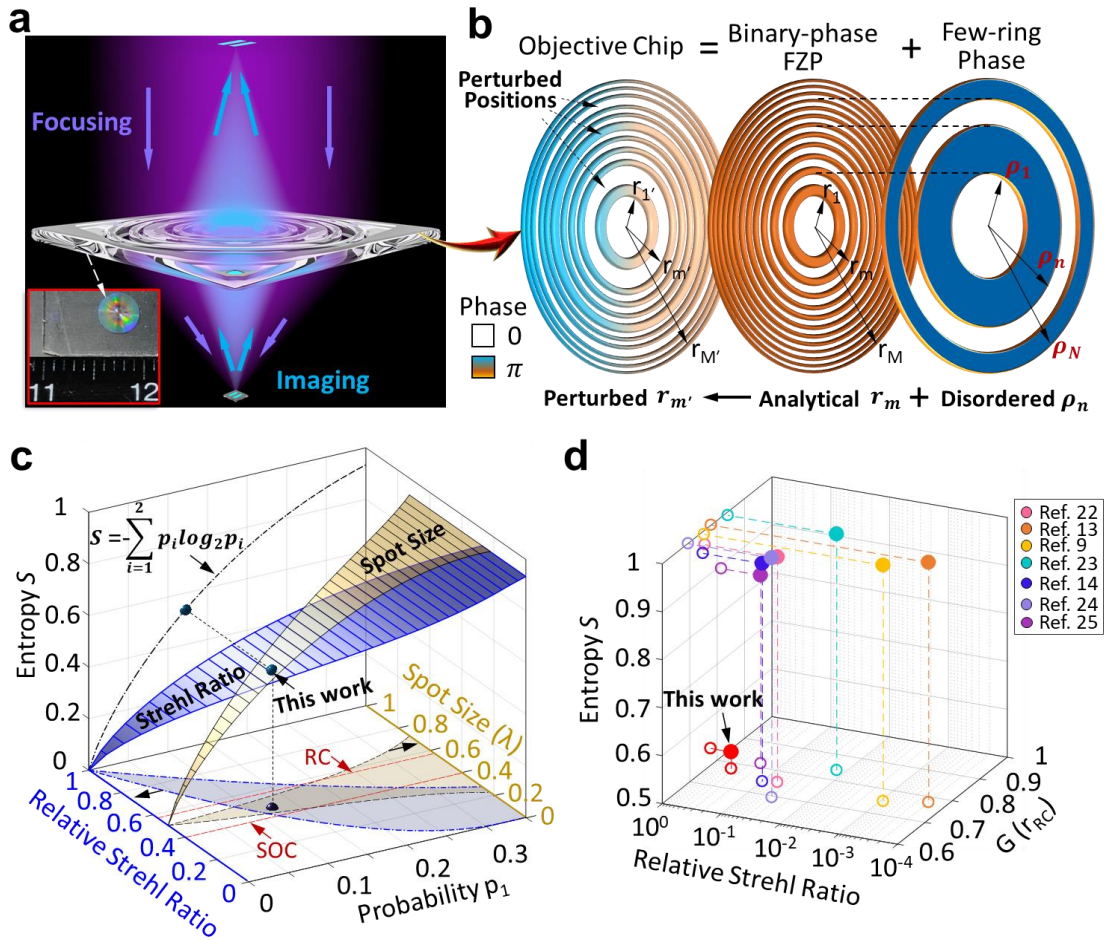

**Figure 1. Working principle of the bifunctional objective chip.** (**a**) Sketch of the objective chip with both focusing and imaging functionalities. (**b**) Design principle of the objective chip composed of a FZP and a few-ring phase with weak disorder. (**c**) Simulated relative Strehl ratio and spot size under different deviation probabilities of $0 \leq p_1 \leq 0.35$. The information entropy $S$ is also provided together as a function of $p_1$. The correlation among the Strehl ratio (blue), $p_1$ and $S$ yields a three-dimensional plotting that visualizes the underlying link between the Strehl ratio and S straightforwardly. Similarly, the relationship between spot size (yellow) and $S$ is also shown with a three-dimensional configuration. Both three-dimensional drawings are projected to the longitudinal plane of $S=0$ for a better observation, where the Rayleigh criterion (RC) and super-oscillation criterion (SOC) are shown in red-dashed lines to distinguish the subdiffraction-limit focusing. (**d**) Entropy $S$, relative Strehl ratio and focal size of other reported planar lenses that provide the structural parameters (which are used to output $p_1$ for calculating entropy $S$) in their publications. For a fair comparison, all the focal sizes are normalized to the Rayleigh diffraction limit of $r_{RC}=0.51\lambda/NA$ (in terms of FWHM). The solid circles denote these three parameters of all these reported lenses (distinguished by colors), while the hollow circles are their projections on different two-dimensional planes for a clear demonstration. Its extended version with more data is provided in Supplementary Fig. 5b for a better observation.

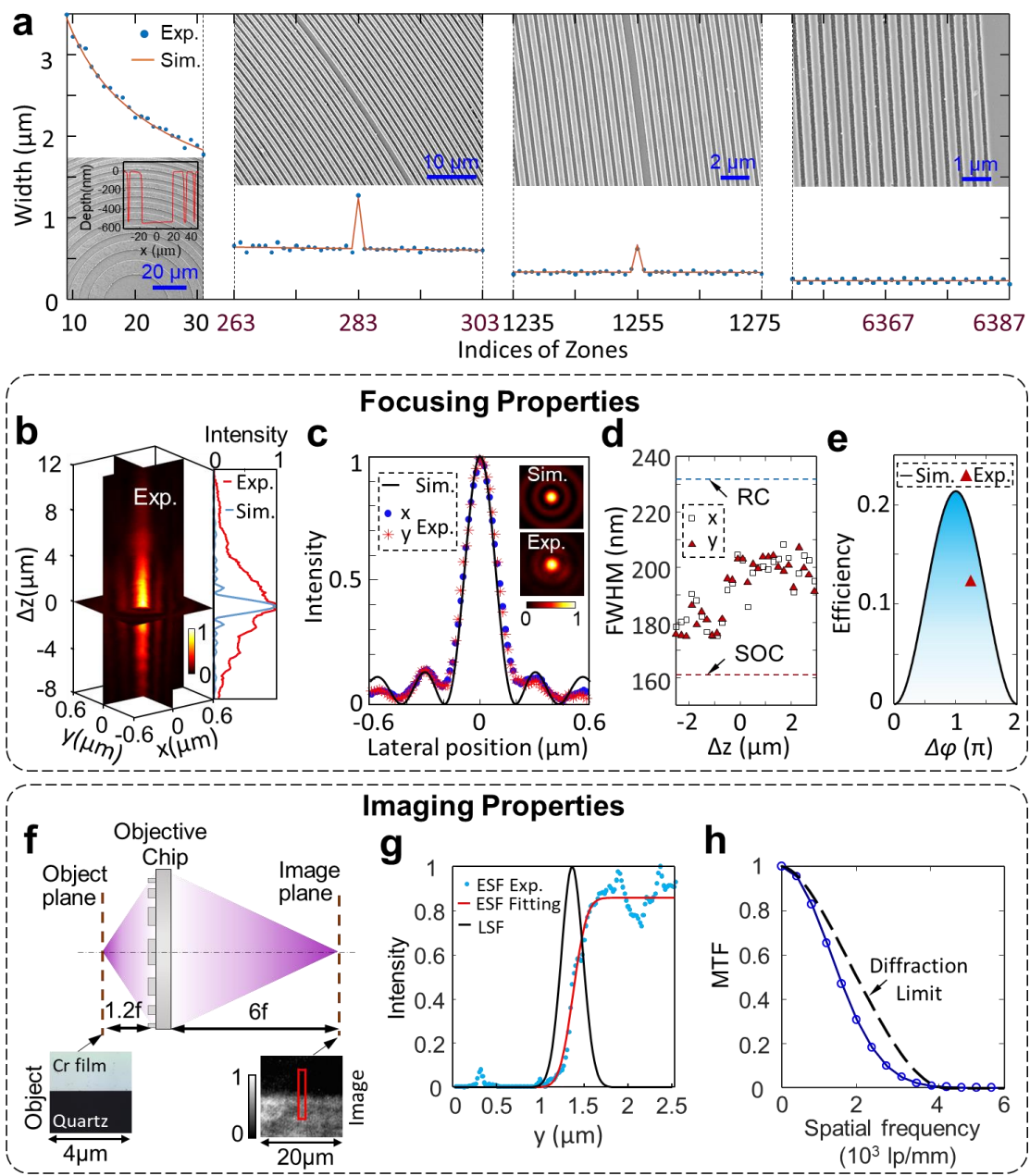

**Figure 2. Focusing and imaging properties of our objective chip.** (a) Simulated (curves) and experimental (dots) widths of belts at the different regions of our objective chip. Inset: SEM images of the different regions by addressing the corresponding zone numbers. The etched depth (left-

bottom) around the center region of the objective chip is measured experimentally by using a profilometer. **(b)** Optical field near the focal plane of the objective chip. Cross sections of the measured intensity profiles are shown in the left panel, while the right panel shows a comparison between the simulated and experimental on-axis line-scanning intensity profiles. **(c)** Simulated (curve) and experimental (dots and asterisks) line-scanning intensity profiles at the focal plane of objective chip. Their 2-dimensional intensity profiles in the region of 1.2 μm×1.2 μm are provided in the insert. **(d)** Lateral (squares for the *x* direction and triangles for the *y* direction) FWHMs of the measured spot near the focal plane (*i.e.*, $\Delta z=z-f=0$). RC: Rayleigh criterion ($0.51\lambda/NA$); SOC: Superoscillation criterion ($0.358\lambda/NA$). **(e)** Simulated (curve) and experimental (triangle) efficiency when the phase difference $\Delta\varphi$ between two neighboring etched and unetched parts changes from 0 to $2\pi$. **(f)** Sketch for wide-field imaging by using our objective chip. The object and image distances are 1.2*f* and 6*f* (the focal length *f*=1mm) respectively, yielding a magnification of 5×. Such a magnification is chosen to avoid optical aberration, while it is enough to demonstrate the capability of collecting light with high spatial frequencies. Inserts: the knife-edge object (left, captured by using a reflective microscope that generates bright chromium film and dark quartz substrate) and its image (right) taken by using our objective chip. **(g)** Experimental edge spread function (ESF, which is calculated by using the average intensity along the long side of the red box in the insert of **(e)**). To evaluate its resolving power, the spatial coordinate *y* is scaled down by its magnification of 5. The experimental ESF is fitted by an error function, the deviation of which outputs the line spread function (LSF). **(h)** Retrieved modulation transfer function (MTF, solid-circle curve) of the objective chip by using the Fourier transform of the achieved LSF in **(g)**. The diffraction limit (dashed curve) is also provided for a better comparison.

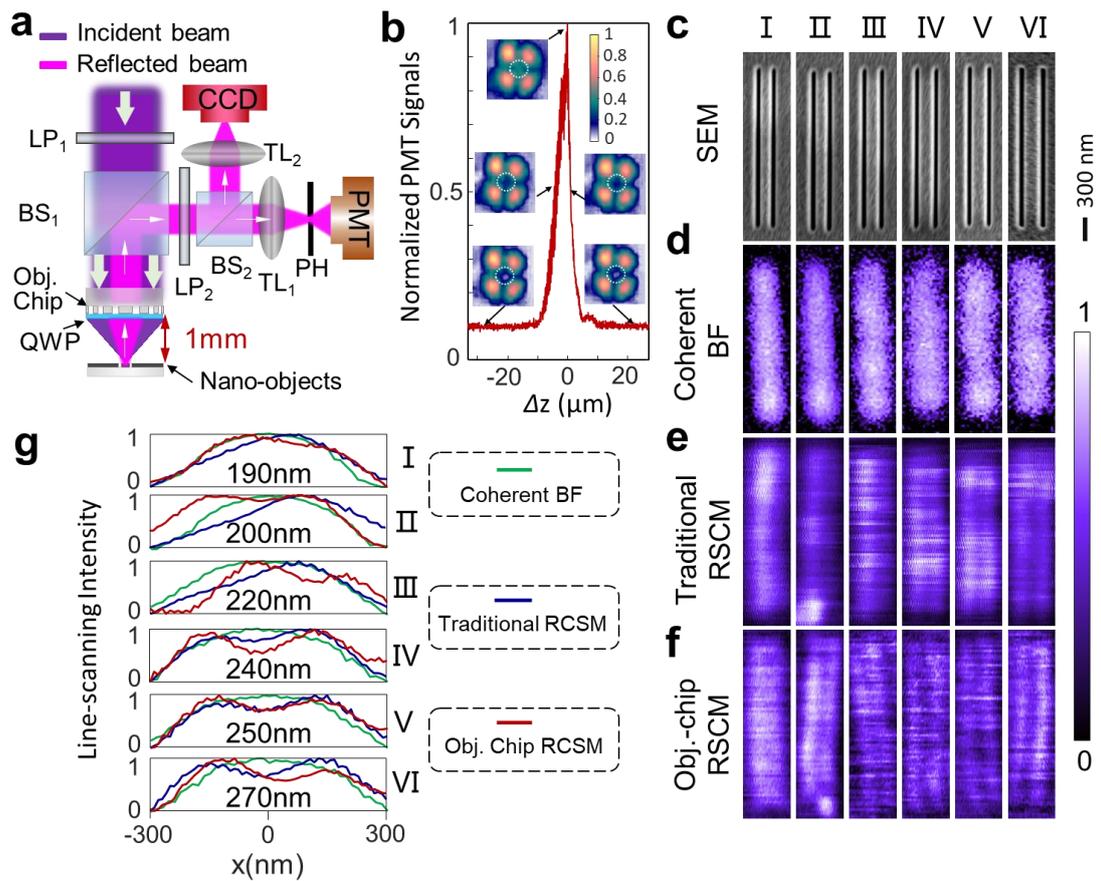

**Figure 3. Reflective scanning confocal microscopy based on our objective chip**. (**a**) Sketch of the optical setup of the objective-chip-based reflective SCM. LP: linear polarizer; BS: beam splitter; QWP: quarter wave plate; L: lens; PH: pinhole; PMT: photomultiplier tube. (**b**) Detected signals (PMT, curve) and images (CCD, inserts) when the nano-objects are scanned with the out-of-focus distance $\Delta z=z-f$. In the CCD images, the dashed circles denote the position of the focused signal light. In the PMT signals, the nonzero background (~0.1) is caused by the incompletely suppressed light (*i.e.*, the four-lobe patterns) reflected from the back surface of the objective chip. (**c-f**) Double slits (**c,** SEM) and their images by using coherent bright-field microscopy (**d**), traditional RSCM (**e**) and objective-chip-based RSCM (**f**). The CTC distances of double slits I to VI are 190 nm, 200 nm, 220 nm, 240 nm, 250 nm and 270 nm, respectively. The height and width of each slit are 2 μm and 50 nm respectively. Scale bar: 300 nm. (**g**) Line-scanning intensity profiles of images by using different microscopies.

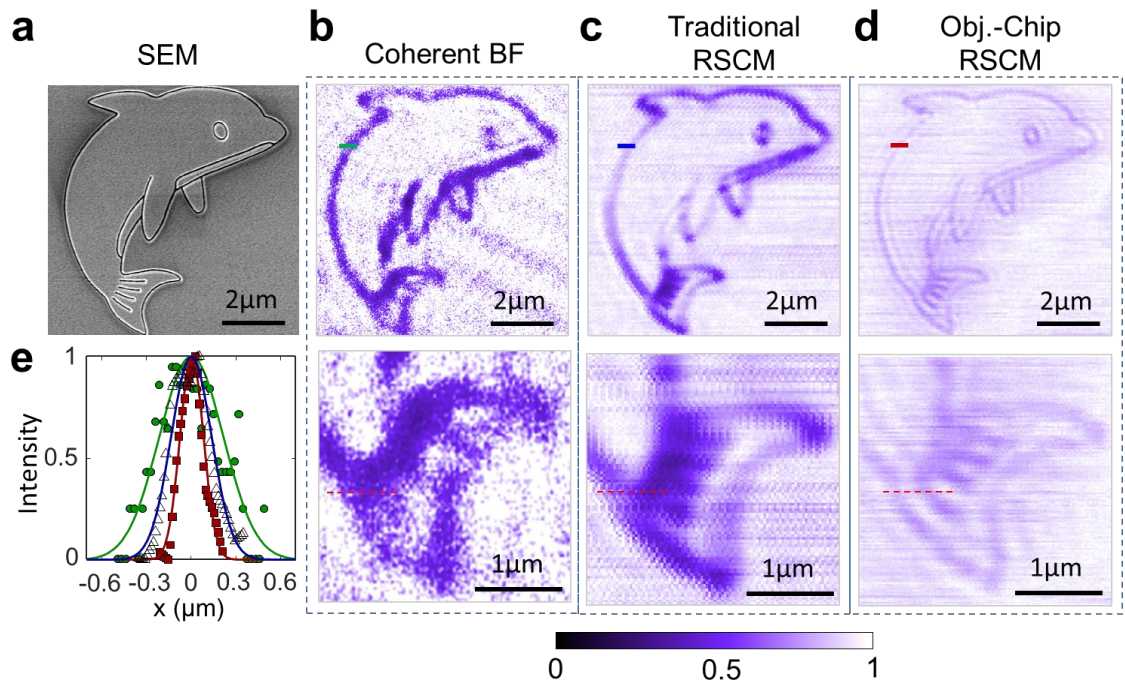

**Figure 4. Imaging complex nano-objects by using different microscopies**. (**a**) SEM image of a "dolphin" object composed of 50 nm-width curves. (**b-d**) Images (upper row) by using coherent bright-field microscopy (**b**), traditional RSCM (**c**) and objective-chip-based RSCM (**d**). Their zoomed-in images with more details are shown in the lower row. (**e**) The line-scanning profiles of the imaged dolphins (along the colored lines of the upper low in **b-d**) by using coherent bright-field microscopy (green circles), traditional RSCM (triangles) and objective-chip-based RSCM (red squares). These line-scanning data are fitted by using a Gaussian shape to guide the eyes.

Supplementary Materials for

# An entropy-controlled objective chip for reflective confocal microscope with sub-diffraction-limit resolution

Jun He[1,#], Dong Zhao[1,#], Hong Liu[2], Jinghua Teng[2,*], Cheng-Wei Qiu[3,*], Kun Huang[1,*]

[1]Department of Optics and Optical Engineering, University of Science and Technology of China, Hefei, Anhui 230026, China

[2]Institute of Materials Research and Engineering, Agency for Science Technology and Research (A*STAR), 2 Fusionopolis Way, #08-03, Innovis, Singapore 138634, Singapore

[3]Department of Electrical and Computer Engineering, National University of Singapore, 4 Engineering Drive 3, Singapore 117576, Singapore

[#] *J. H. and D. Z.* contributed equally to this work.

[*]Corresponding authors: K. H. (huangk17@ustc.edu.cn), J. T. (jh-teng@imre.a-star.edu.sg) or C. Q. (chengwei.qiu@nus.edu.sg)

## Table of Contents

**Section 1. Mathematical basis for Strehl ratio and focal size of an objective chip**

To reveal the relationship between the information entropy $S$ and optical properties of an objective chip, we investigate its Strehl ratio and focal size under different deviation probability $p_1$. For a given $p_1$, the relative Strehl ratio and focal size change because the locations of the deviated zones in the objective chip are different. It means that the relative Strehl ratio and focal size have a certain range with the fixed minimum and maximum values, which can be determined mathematically by using diffraction properties of each zone.

First, we derive the minimum and maximum Strehl ratios. Because the binary-phase objective chip is reported here with the modulation phase of 0 and $\pi$, we can directly use its complex modulation of 1 and -1, respectively. In our design strategy, the objective chip is functionally divided into a binary-phase FZP and an $N$-ring phase mask. The phase of $\pi$ in the $N$-ring phase mask realizes the reversal (from 1 to -1, or from -1 to 1) of the complex modulation. Assuming that the odd and even rings have the phase of 0 and $\pi$ respectively, it means that the electric fields contributed by the zones in the even rings are removed from those of the standard FZP and then are used to interfere constructively with those of the zones in the odd rings. For the objective chip containing a $N$-ring phase mask, Eq. (3) describing the total electric fields of our objective chip can be rewritten as

$$E_{chip}(u,v,z) = \sum_{n=0}^{N-1}(-1)^n \left[\sum_{m=M_n}^{m=M_{n+1}}(-1)^m A_m\right] = \sum_{m=1}^{m=M}(-1)^m A_m - 2\times\sum_{m\in\mathcal{R}}(-1)^m A_m$$
$$= E_{FZP}(u,v,z) - 2\times\sum_{m\in\mathcal{R}}(-1)^m A_m, \tag{S1}$$

where $\mathcal{R}$ denotes a set of the indices of all the phase-reversed zones (*i.e.*, contained in the even rings of the $N$-ring phase mask). For our design strategy used in this work, we can obtain the number of the set $\mathcal{R}$ by using $M\cdot p_1$, where $M$ is the total number of zones in the corresponding FZP and $p_1$ is the deviation probability (see Eq. (1) in the main text) of the phase-reversed zones. Considering its university, Eq. (S1) is valid for all binary-phase planar lenses.

After substituting Eq. (S1) into Eq. (2) of main text, the relative Strehl ratio can be expressed as

$$SR = \frac{I_{chip}(0,0,z=f)}{I_{FZP}(0,0,z=f)} = \frac{|E_{chip}(0,0,z=f)|^2}{|E_{FZP}(0,0,z=f)|^2} = \frac{|E_{FZP}(0,0,f)-2\times\sum_{m\in\mathcal{R}}(-1)^m\cdot A_m(0,0,f)|^2}{|E_{FZP}(0,0,z=f)|^2}, \tag{S2}$$

where $E_{FZP}(0, 0, f)$ is in phase with $a_0=(-1)^m A_m(0, 0, f)$. Meanwhile, for different $m$, the item $a_0$ is also in phase with each other and nearly a constant with a slow variation from $a_{0\min}=0.87$ to $a_{0\max}=2$, which can be numerically calculated by using the rigorous Rayleigh-Sommerfeld diffraction integral. Due to their feature of slow variation, the on-axis intensity $a_0$ from all the zones with the zone indices $m \in \mathcal{R}$ are assumed to be identical. Thus, $\sum_{m \in \mathcal{R}} a_0$ can be approximated by $a_0 p_1 M$. By applying these assumptions, Eq. (S2) can be approximated as

$$SR = (1 - 2a_0 p_1 M / \sqrt{I_{FZP}})^2, \tag{S3}$$

where $E_{FZP} = \sqrt{I_{FZP}}$ and $0.87 \leq a_0 \leq 2$. Eq. (S3) reveals the direct link between SR and $p_1$ (which is a key parameter to evaluate the disorder of a planar lens). Based on it, we can evaluate the range of SR by determining its minimum and maximum values. By carrying out straightforward derivations, we have the minimum value $SR_{min} = (1 - 2a_{0max} p_1 M / \sqrt{I_{FZP}})^2$ for $0 \leq p_1 \leq \sqrt{I_{FZP}}/(2Ma_{0max})$ and $SR_{min} = 0$ for $\sqrt{I_{FZP}}/(2Ma_{0max}) \leq p_1 \leq 0.5$, and the maximum value $SR_{max} = (1 - 2a_{0min} p_1 M / \sqrt{I_{FZP}})^2$ for $0 \leq p_1 \leq \sqrt{I_{FZP}}/(Ma_{0max} + Ma_{0min})$ and $SR_{max} = (1 - 2a_{0max} p_1 M / \sqrt{I_{FZP}})^2$ for $\sqrt{I_{FZP}}/(Ma_{0max} + Ma_{0min}) \leq p_1 \leq 0.5$. Note that the analytical $SR_{min}$ and $SR_{max}$ depend on only the deviation probability $p_1$, and are therefore valid for various objective chips with different $N$.

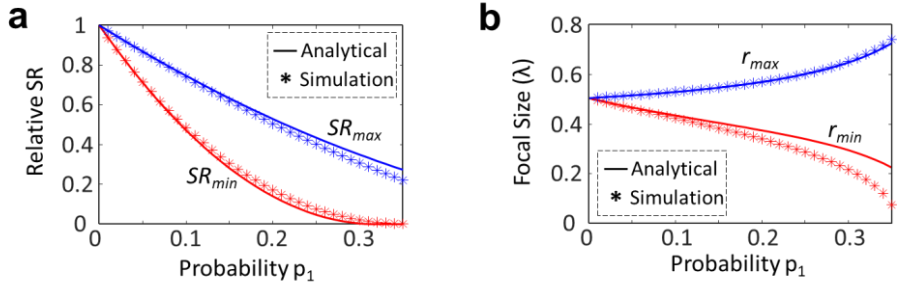

**Supplementary Fig. 1. Analytical and simulated Strehl ratio (a) and focal size (b) under the different deviation probability $p_1$.** The analytical Strehl ratios are determined by using Eq. (S3), which the analytical focal sizes are obtained by numerically solving Eqs. (S4) and (S5).

To verify the analytical $SR_{min}$ and $SR_{max}$, we have simulated the range of SR by using the proposed objective chip with a 5-ring phase mask. The limited $N=5$ of the phase mask allows us to go through all possible solutions quickly without any optimization because all the $A_m$ can be calculated ahead. By controlling the number and position of the phase-reversed zones in the 5-few ring phase mask, the simulations are implemented within the range of $0 \leq p_1 \leq 0.35$ with an interval of 0.01, which is enough here because low disorder $p_1$ is important to develop the objective chip with good balance between imaging and super-focusing. The simulated $SR_{min}$ and $SR_{max}$ are shown in Supplementary Fig. S1a, exhibiting good agreement with their analytical values. Their slight deviations come from the approximations made during its derivation. Therefore, these results have confirmed that the analytical $SR_{min}$ and $SR_{max}$ give a good prediction for the range of Strehl ratio.

Second, the focal size of an objective chip with different $p_1$ can also be predicted by using Eq. (S1). As shown in Eq. (S1), the electric field of the objective chip is taken as the coherent superposition of the electric fields from all the zones. Diffraction behavior of each zone is important in predicting the focal size of the objective chip. Because the width of each zone is small, the diffraction field from each zone is mainly determined by its focusing angle between the outmost boundary of each zone and optical axis. When the focusing angle is large, the relative focal spot size of diffraction field from one zone is small; vice versa. For an objective chip, its maximum focal spot is achieved when the contribution from the outermost zones is small, where the electric field at the focal plane is

$$E_{chip}(u,v,f) = E_{FZP}(u,v,z) - 2 \times \sum_{m=M\cdot(1-p_1)}^{m=M}(-1)^m A_m, \qquad (S4)$$

where the phases of the outermost zones are reversed with a deviation probability $p_1$. From Eq. (S4), we numerically predict the maximum focal size, which depends on only $p_1$. In contrast, when the phases of the inner zones are reversed, the minimum focal size can be predicted by using the electric field at the focal plane

$$E_{chip}(u,v,f) = E_{FZP}(u,v,z) - 2 \times \sum_{m=0}^{m=M\cdot p_1}(-1)^m A_m. \qquad (S5)$$

By using Eqs. (S4) and (S5), we can calculate the maximum and minimum focal spots, as shown in the solid lines in Supplementary Fig. S1b. Because both Eqs. (S4) and (S5) have no limitation about the number $N$, we use our proposed objective chip with a 5-ring phase mask to verify the predicted minimum ($r_{min}$) and maximum ($r_{max}$) focal size. Similarly, we go through all the possible solutions by changing the number and position of the phase-reversed zone in the second and fourth rings of the few-ring phase mask, which can be implemented together with the above calculation of the Strehl ratio. The simulated $r_{min}$ and $r_{max}$ are provided in Supplementary Fig. 1b. Both the predicted and simulated $r_{max}$ agree with each other for the interested range of $0 \leq p_1 \leq 0.35$. However, for the case of $r_{min}$, the discrepancy between the prediction and simulation increases with the increment of $p_1$, which is caused by the large error of the approximation in Eq. (S5). In fact, for a large $p_1$, the disorder in the objective chip is higher than the predicted one in Eq. (S5), so that the minimum focal spot has more choices with a larger range than that predicted by Eq. (S5). Despite this, our prediction in Eq. (S5) shows the same decreasing tendency when $p_1$ increases, thereby confirming the validity of the predicted focal sizes.

**Section 2. Optimization of the objective chip**

According to the design strategy described in the main text, we implement the optimization of the objective chip with four key steps, as discussed below.

2.1. Determining the radius of each belt in a standard zone plate

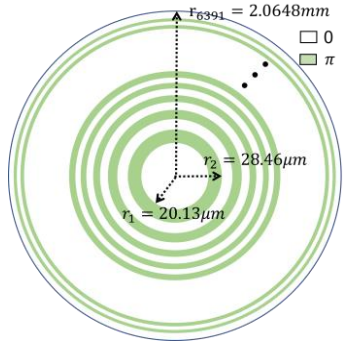

**Supplementary Fig. 2. Sketch of standard zone plate with 6391 belts.** The imaging parameters are λ=405 nm, $f$=250 μm, which indicates a radius of 2 mm.

Considering the non-paraxial feature of this objective chip, a rigorous formula of standard binary phase Fresnel zone plate (BPFZP) should be used to calculate the radius $r_m$ of the $m^{th}$ belt with

$$r_m = \sqrt{(\tfrac{\lambda}{2}m)^2 + m\lambda f}, \tag{S6}$$

where $m$=0, 1, 2,…, $M$, the wavelength λ=405 nm, the focal length $f$ =1mm. In our design, the total number $M$ of belts in this BPFZP is $M$=6391 and the radius of BPFZP is ~2.0648mm, which yields a numerical aperture (*NA*) of 0.9, as shown in Supplementary Fig. 2. The binary phase is employed to enhance optical efficiency of the objective chip.

2.2. Calculating the focal field of each belt in the BPFZP

Benefiting from our design strategy, no new ring is created during our optimization because all the radii can be described by Eq. (S6). It means that the focal field of light from each zone can be calculated ahead of optimization and then stored in a database, so that we directly revisit the relative focal field during the optimization. Thus, the time cost will be significantly shorten. Since we need to evaluate the lateral focal size and the longitudinal depth of focus, both focal fields along the radial (*i.e.*, *r*) and longitudinal (*i.e.*, *z*) direction are calculated ahead, where the positons of interest are: 1) the lateral positions 0≤*r*≤λ at the focal plane *z*=1000 μm; 2) the longitudinal positions 950 μm≤*z*≤1050 μm at the on-axis position *r*=0. By using Rayleigh-Sommerfeld diffraction theory without any approximation for high accuracy, we calculate the focal fields of each belt along the lateral and longitudinal positions, and store them in two matrices (*i.e.*, $A_r$ and $A_z$, see Supplementary Figs. 3a and 3b) respectively. According to the electric field stored in the 1$^{th}$, 3000$^{th}$ and 6391$^{th}$ column of $A_r$ and $A_z$, the normalized intensity of diffraction field at two target positions of corresponding belts of BPFZP are exemplified in Supplementary Figs. 3c and 3d, respectively. To show its convenience, the focal fields at two target positions for a BPFZP can be calculated as

$$E(r, z = f) = \sum_{m=0}^{m=M}(-1)^m A_m(r),$$

$$E(r = 0, z) = \sum_{m=0}^{m=M}(-1)^m A_m(z) , \qquad (S7)$$

where $A_m(r)$ is the radial-position electric field of light diffracting from the *m*$^{th}$ belt in the zone plate and saved in the *m*$^{th}$ column of $A_r$, $A_m(z)$ is the longitudinal-position electric field of light diffracting from the *m*$^{th}$ belt in the zone plate and stored in the *m*$^{th}$ column of $A_z$. According to this definition, $A_m(r)=A_m(z)=0$ when *m*=0. Based on these database, we can calculate any focal field of the objective chip only if the phase of each zone is given.

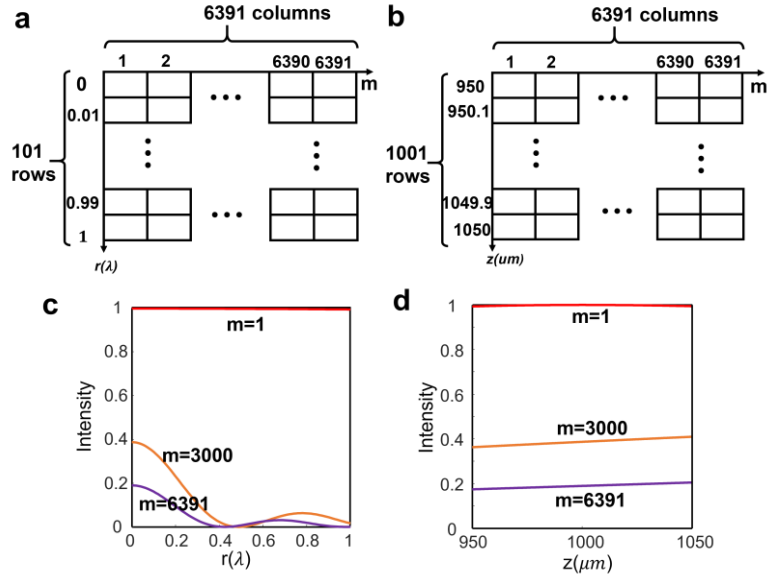

**Supplementary Fig. 3. Data preparation for optimization**. The $A_r$ (**a**) and $A_z$ (**b**) database show

the way of saving the pre-calculated data in our optimization. The normalized line intensity at the focal plane **(c)** and on the optical axis **(d)** of the 1$^{th}$, 3000$^{th}$ and 6391$^{th}$ belt of BPFZP.

2.3. Optimizing the 5-ring phase mask

To optimize the detailed structures of the 5-ring phase mask, we use the well-matured particle swarm optimization (PSO) algorithms that have been used frequently to design various lenses, especially for super-oscillation and super-critical lenses. Considering the limited number of the few-ring mask, the optimization will be implemented with its standard version of the PSO algorithm. For a 5-ring mask, the dimension of particle D is 5. In our algorithm, the size or population of the particle is 20. The details and flowchart of PSO are shown in Supplementary Fig. 4. Since the electric fields ($A_r$ and $A_z$) have been calculated in advance, the calculation of fitness of each particle in iteration can be finished quickly, as shown below.

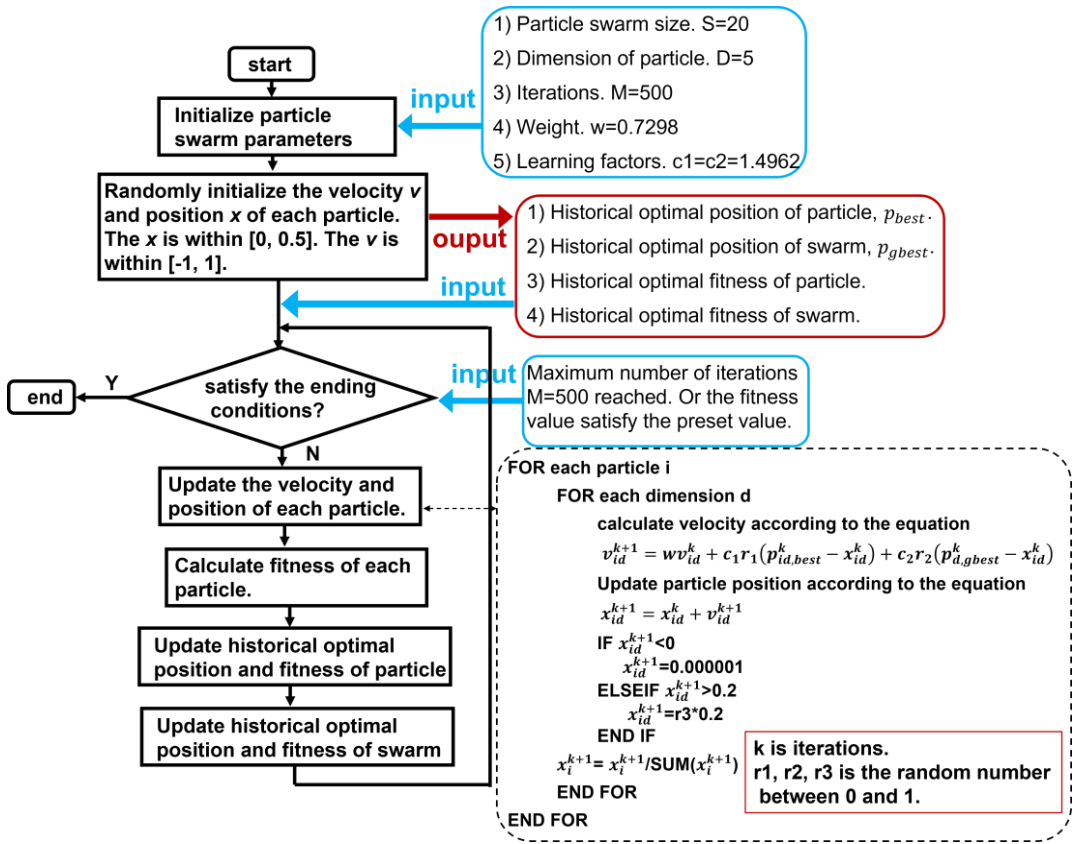

**Supplementary Fig. 4. The detailed flowchart of our built PSO.**

To correlate the particle parameters $[x_1, x_2, x_3, x_4, x_5]$ with the unknown structures of the designed objective chip, we define the relative *NA* at the each boundary of the 5-ring phase mask by using

$$NA_n = \frac{\sum_{i=1}^{n} x_i}{\sum_{i=1}^{5} x_i} \cdot NA_0, \tag{S8}$$

where $n$ is the ring number of the few-ring phase mask, $NA_0$=0.9 is used in this work. From Eq. (S8), one can induce that $NA_5=NA_0$, which means that the outer boundary of the 5-th ring in the phase mask refers to the maximum radius of 2.046 mm. The five parameters in each particle are related with the difference of $NA$ between two neighboring rings. In this definition, we have built the one-to-one relationship between the particle parameters and the structures of the objective chip. In addition, such a definition will offer full degree of freedom to go through all the possible solutions because each $x_i$ can be valued within $0<x_i<\infty$. The normalization factor of $1/\sum_{i=1}^{5} x_i$ is quite helpful to make the maximum value of $NA_0$, thus enabling each particle to yield a physically meaningful objective chip. Thus, the universal properties of design an objective chip are maintained in our definition, which is an important step to implement this optimization. In one iteration, each $x_i$ is updated with the PSO algorithm (as described in the dashed rectangle of Supplementary Fig. 4). Based on the optimized $x_i$, we derive the $NA$ parameters in Eq. (S8), from which we find each $R_n$ in the designed objective chip by using the equation $R_n/\sqrt{R_n^2 + f^2} = NA_n$ (or $R_n = NA_n \cdot f/\sqrt{1 - NA_n^2}$).

To avoid the creation of additional finer structures when combining the zone plate and 5-ring mask, the optimized $R_n$ is approximated by the closet $r_m$, which is labelled as $r_{M_n} = \rho_n$. Thus, we can find all the $M_n$, hereby fixing the geometric structures of the few-ring phase mask in each iteration.

According to the updated $M_n$ and the well-built database, we can calculate the electric fields of objective chip at two positions of interest:

$$E(r, z = f) = \sum_{n=0}^{n=4} \sum_{m=M_n}^{m=M_{n+1}+1} (-1)^{m+n} A_m(r),$$

$$E(r = 0, z) = \sum_{n=0}^{n=4} \sum_{m=M_n}^{m=M_{n+1}} (-1)^{m+n} A_m(z). \tag{S9}$$

From Eq. (S9), we can obtain the relative intensity $I_r = |E(r, z = f)|^2$ and $I_z = |E(r = 0, z)|^2$. Then, two root-mean-square error (RMSE) between ideal and simulated patterns are calculated as

$$RMSE_1 = \sqrt{\frac{(I_r - I_r^{ideal})^2}{N_r}}$$

$$RMSE_2 = \sqrt{\frac{(I_z - I_z^{ideal})^2}{N_z}}, \tag{S10}$$

where the ideal radial-position intensity $I_r^{ideal}=|J_0(krNA)|^2$, $J_0$ is the zero-order Bessel function of the first kind, $k=2\pi/\lambda$ is wave vector, the radial position $r$ is valued between 0 and $\lambda$ with an interval of $0.01\lambda$ (*i.e.*, the sampling number is $N_r=101$), the ideal longitudinal-position intensity $I_z^{ideal} = e^{-\frac{(z-f)^2}{DOF^2}}$, $DOF = \frac{\lambda}{1-\sqrt{cos\theta}}$, $sin\theta = NA_0 = 0.9$, the longitudinal position $z$ is valued between 950 μm and 1050 μm with an sampling interval of 0. 1 μm (*i.e.*, the sampling number is $N_r=1001$). The RMSE$_1$ and RMSE$_2$ are used to evaluate the electric fields at the radial and longitudinal positions. Based on them, we can build the cost function by using $CF=C\cdot RMSE_1 + RMSE_2$, where the positive constant $C$ can be adjusted according to any special requirement. In this work, $C=2$ is used to realize the sub-diffraction-limit focusing.

After 500 iterations in the PSO algorithm, we finally obtain the $M_1=283$, $M_2=850$, $M_3=1046$, $M_4=1258$ and $M_5=6391$, and the additional phase mask can be sketched in Supplementary Fig. 5. The combination between standard zone plate (Supplementary Fig. 2) and additional phase masks (Supplementary Fig. 5a) forms the objective chip. The simulated intensity profiles at the longitudinal and radial positions are provided in Figs. 2b and 2c of main text, respectively.

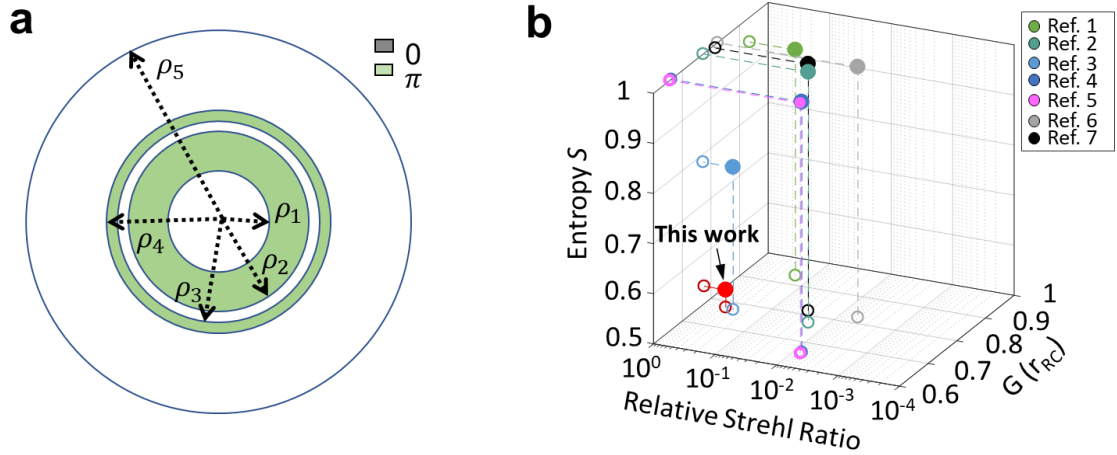

**Supplementary Fig. 5. Design results of a 5-ring phase mask.** **(a)** Phase profile of the optimized few-ring mask. The ideal phase difference $\Delta\varphi$ between two neighboring rings is π. **(b)** Entropy $S$, relative Strehl ratio and focal size of other reported planar lenses with the structural parameters (which is used to output $p_1$ for calculating entropy $S$) provided in their corresponding publications. This figure is an extension of Fig. 1d in the main text for a better observation due to the overlay of data.

To show the difference from the previous planar lenses, we provide the entropy, relative Strehl ratio and focal size of various reported lenses[1-7] in Supplementary Fig. 5b, which is an extension of Fig. 1d in the main text for a better observation. From both figures, we can conclude that the entropy S of our proposed objective chip is the closet to the equilibrium point $S_0=0.5$, implying the good balance between imaging and super-focusing. Note that, although Ref. 3 has nearly identical focal size and relative Strehl ratio to ours, its entropy *S* is much higher than *S*=0.535 in our objective chip, implying that its imaging capability is poor. For the other lenses, their entropy approaches 1, which is a natural result when a lens is designed only for super-focusing with high disorder. Therefore, the information entropy is a good measurement of evaluating the imaging and super-focusing capabilities in a straightforward way.

**Section 3. Experimental characterization of focusing properties of the objective chip**

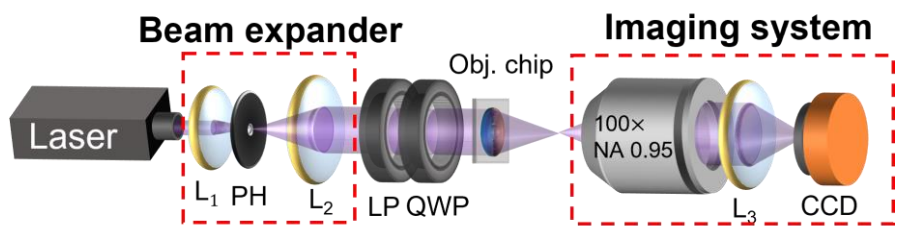

**Supplementary Fig. 6. Experimental setup for characterizing the focusing capability of the fabricated objective chip.** The beam expander consists of lenses $L_1$ and $L_2$ (with the focal lengths of 25.4 mm and 300 mm respectively) and a 25μm-diameter pinhole. The imaging system consists of 100× objective with 0.95 NA, a lens $L_3$ (with a focal length 400 mm) and a CCD camera. LP: linear polarizer. QWP: quarter-wave plate; Obj. chip: objective chip;

To characterize the focusing capability of objective chip, we use the experimental setup as shown in Supplementary Fig. 6. The beam from a λ=405 nm laser is reshaped by a beam expander (consisting of $L_1$, $L_2$, and a 25 μm-diameter pinhole), yielding a fundamental Gaussian beam with the diameter of ~1 mm. To generate circularly polarized illumination, a linear polarizer and quarter-wave plate are employed for obtaining the highly axisymmetric focal spot. Then, a sub-diffraction-limit focal spot generated by objective chip is projected by using the imaging system (composed of a 100× objective with 0.95 *NA* and a lens $L_3$) onto the CCD camera.

The focusing process of this objective chip is recorded dynamically in Supplementary Video 1, which presents the focused spot near the focal plane. Such a video shows the same intensity profiles that are similar to those in Fig. 2b of main text. All these experimental results have confirmed the good focusing with our proposed objective chip.

**Section 4. Measuring the focusing efficiency of the objective chip**

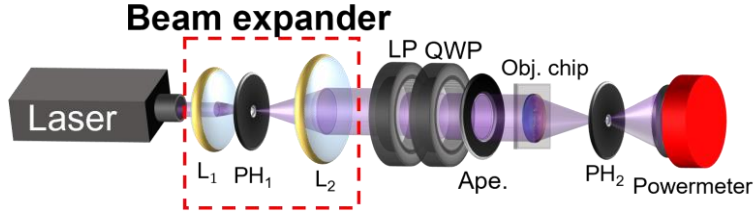

**Supplementary Fig. 7. The schematic diagram of the experimental setup for measuring the focusing efficiency of the objective chip.** The beam expander is composed of two lenses $L_1$ and $L_2$ (with their focal lengths of 25.4 mm and 300 mm, respectively) and a 25 μm-diameter pinhole ($PH_1$). LP: linear polarizer; QWP: quarter-wave plate; Ape.: aperture; Obj. chip: objective chip; $PH_2$: pinhole with 150 μm diameter.

The focusing efficiency of the objective chip is measured by using the experimental setup sketched in Supplementary Fig. 7. To obtain quasi-plane-wave illumination, we reshape a λ=405 nm laser into a ~1 mm-diameter fundamental Gaussian beam by using a beam expander (containing $L_1$, $L_2$, and a 25 μm-diameter pinhole $PH_1$). A circular polarizer composed of a linear polarizer (LP) and a quarter-wave plate (QWP) is used to convert the polarization of the incident beam into the circular polarization, which is helpful to achieve the circular focal spot under the high-NA focusing condition. To remove the background light from high-diffraction-order rings diffracted by the pinhole $PH_1$, an iris aperture with its transmission area slight larger than the entrance of the objective chip is utilized here to keep the nearly uniform incident light. The second pinhole $PH_2$ with a 150 μm diameter is placed at focal plane of objective chip to select the focused power, which is recorded as $I_1$ by using a power meter. Due to the high NA of 0.9 in the objective chip, the divergence angle of light passed through $PH_2$ is large. Therefore, the power-meter is located close to the $PH_2$ for complete collection of all focused light. By removing the objective chip and the $PH_2$ simultaneously, we can measure the total power of the incident beam, as recorded as $I_0$. Finally, we obtain the experimental focusing efficiency of objective chip, *i.e.*, $\eta = \frac{I_1}{I_2} = 12.3\%$, which is tightly consistent with the theoretical 16.9%. Although such an experimental efficiency is not so high compared with the dielectric metasurfaces, we will show its capability in optical imaging and focusing for high-quality confocal microscopy.

**Section 5. Measuring modulation transfer function (MTF) of the objective chip**

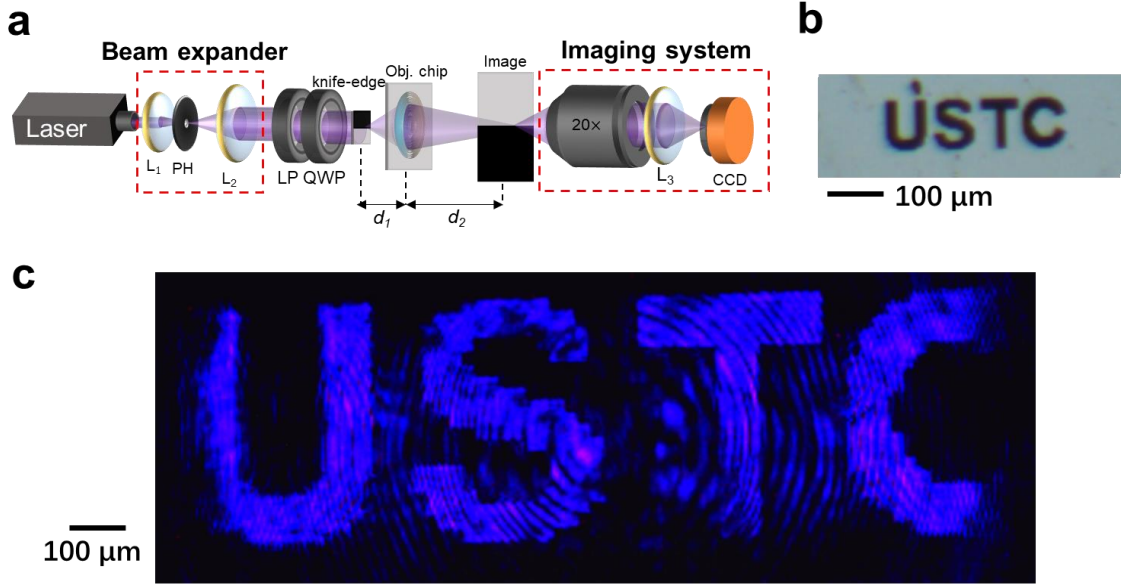

**Supplementary Fig. 8. Sketch for the experimental setup to measure MTF of the objective chip when working in an imaging mode**. (**a**) The beam expander is made of two lenses $L_1$ and $L_2$ (with their focal lengths of 25.4 mm and 300 mm respectively) and a 25 μm-diameter pinhole (PH). The imaging system consists of a 20× objective, a lens (with a focal length of 400 mm) and a CCD camera. LP: linear polarizer; QWP: quarter-wave plate; Obj. chip: objective chip; $d_1$ and $d_2$ are the object and image distances, respectively. (**b-c**) Experimental measurement of field of view by using a large object "USTC" (**b**) with its horizontal length of 310 μm. The relative image is shown in (**c**).

To measure the MTF of the objective chip, a self-made setup presented in Supplementary Fig. 8 is implemented to demonstrate its imaging properties in a transmission mode for a better experimental operation. Note that, the configuration of transmission or reflection will not influence the imaging properties of the objective chip. Similarly, a fundamental Gaussian beam with circular polarization is obtained by using a beam expander and a circular polarizer, and then works as the illumination beam of a knife-edge object (a 140 nm-thick Cr film coated on quartz substrate), which is mounted on 3-dimensional piezo stage (PI). To directly image the knife-edge, the objective chip mounted on a mechanical stage is placed behind the knife-edge with its structure side closer to the knife-edge for the collection of the transmitted light. To adjust the object distance $d_1$, we move the imaging system to see the knife-edge and the structure surface of the objective chip respectively and record their corresponding positions $z_1$ and $z_2$. Thus, the object distance can be evaluated roughly by using $d_1 = |z_1 - z_2|$. We tune the position of objective chip so that the object distance $d_1$=1.2$f$ ($f$=1 mm is the focal length of objective chip) is achieved, which yielding the relative imaging distance of $d_2$=6$f$ (see Fig. 2f in the main text). The captured image is shown in the insert of Fig. 2f in the main text. To show its imaging process of the knife edge, we provide a dynamic video (see Supplementary Video 2) that records the out-of-focus and in-focus images by moving the axial position of knife-edge near $d_1$=1.2$f$ and simultaneously fixing the position of the imaging system. These experimental results clearly show the good imaging ability of our proposed objective chip.

To test its field of view, we use a larger object of "USTC" that has a horizontal length of 310 μm, see Supplementary Fig. 8b. The object "USTC" is placed at z=1.2$f$, which means its magnification of 5X. The resulting image is shown in Supplementary Fig. 8c, which shows the clear image with the slightly blurred horizontal edges. It means that a bigger object with its dimension larger than 310 μm cannot be imaged with clear edges. This indicates that the field of view is 310 μm×310 μm at the magnification of 5X. These results show the wide-field image capability of our developed objective chip.

**Section 6. Roles of collection objective in scanning confocal microscopy**

To highlight the importance of the collection objective in a scanning confocal microscopy, we implement the numerical simulation of the imaging processes by using the theory of scanning confocal microscopy [8]. When the scanned object is not an infinitesimal point, the imaging resolution of scanning confocal microscopy is determined by the *NA* of collection objective. In our simulations, the theoretically focused spot by using our objective chip is taken as the focal field $h_1$ of the condenser lens while the PSF of collection objective is the well-known Airy spot $h_2 = \frac{J_1(krNA)}{krNA}$, where *NA* is the numerical aperture the collection objective, $k=2\pi/\lambda$ is the wave number, λ is the wavelength and $r$ is the radial coordinate. The transmission *T* of the nano-objects is taken as its original pattern without considering the light-structure interaction, for the simplicity of the entire simulation process. For a certain scanning position ($x_s$, $y_s$), the electric field at the pinhole plane (*i.e.*, the imaging plane of the collection objective) can be written as $(h_1 \cdot T) \otimes h_2$, where $\otimes$ stands for the convolution operation. After selected by the pinhole with a circular aperture, the total power is taken as the value of image at this scanning position ($x_s$, $y_s$). Thus, if all the positions in the object are scanned, we can achieve the final image. All the codes are built in a Matlab software. Supplementary Fig. 8 show the simulated images by using collection objectives with different *NA*s that changes from 0.5 to 0.9. From these simulated results, one can see that, when *NA*≤0.6, the scanning images will be distorted seriously with a bad imaging resolution. In comparison, the imaging resolution is high if *NA*≥0.7, which suggests the smallest *NA* of 0.7 to achieve high-resolution imaging in a SCM. Therefore, the *NA* of collection objective is also critical for high-quality imaging of real objects in SCM. Moreover, the resolution reduces with the increment of slit width, because the contrast of image is getting lower. This result is consistent with the fact that the fixed 200nm-CTC-distance double slits will not be resolved when the double silt merges into one slit with the continuing increment of slit width.

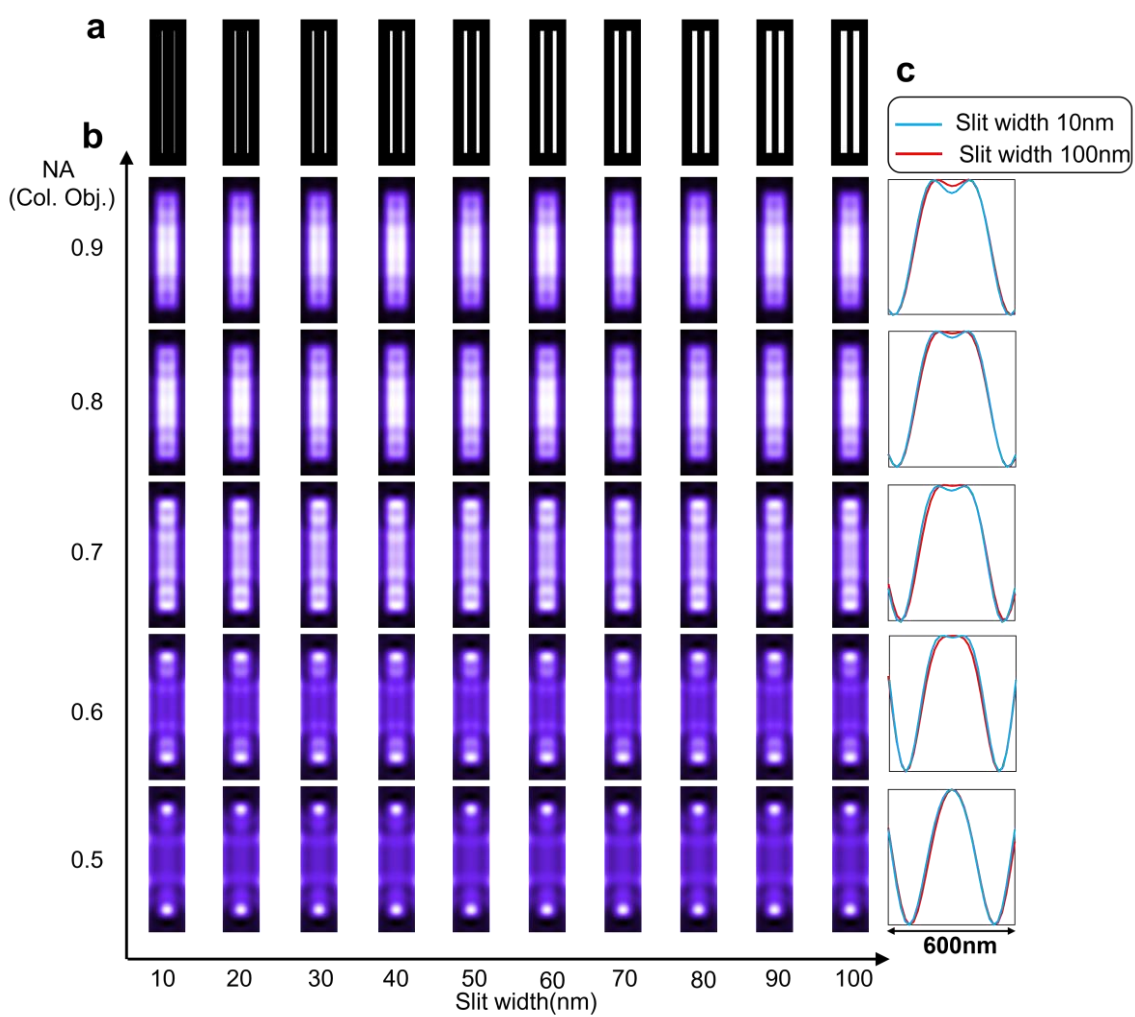

**Supplementary Fig. 9. Simulated images of double slits by using collection objectives with different *NA*s. (a)** Double slits with the center-to-center distances of 200 nm and the silt widths ranging from 10nm to 100nm. The length of slits are 2 μm. **(b)** Simulated scanning images under different *NA*s (0.5-0.9) and slit widths (10nm-100nm). **(c)** The line-scanning profiles of these slit images for the collection objectives with different *NA*s.

**Section 7. Simulated images by using different microscopies for comparison**

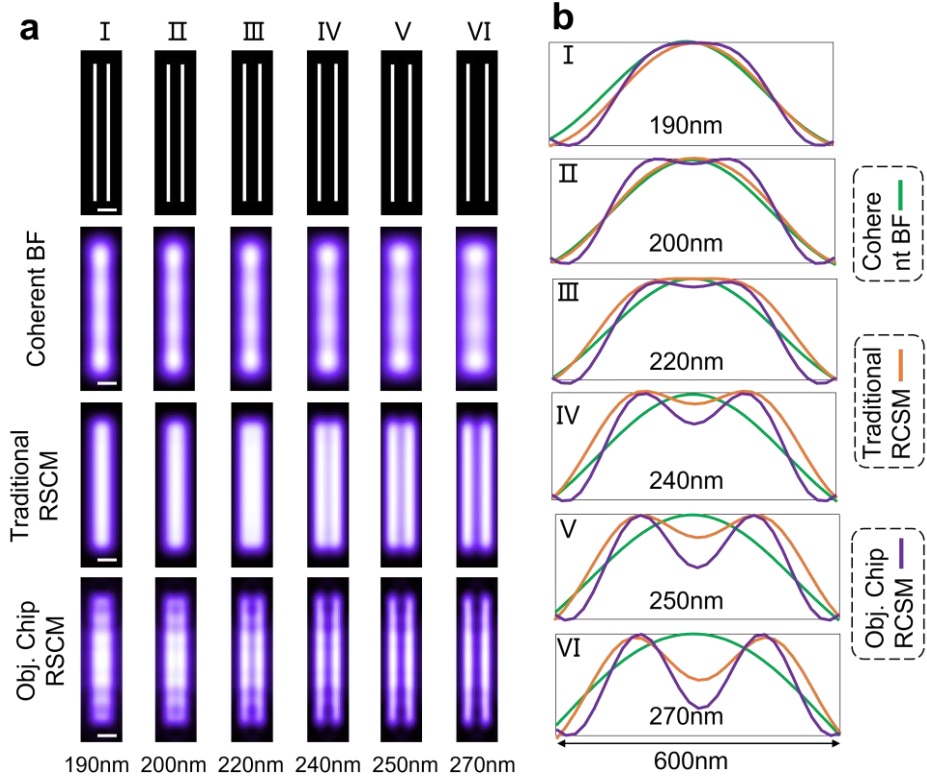

**Supplementary Fig. 10. Simulated images of double slits with different CTC distances. (a)** Sketch of double slits (top row) and their simulated images by using coherent bright field microscope (second row), traditional RSCM (third row) and objective-chip-based RSCM (forth row). The height and width of slit are 2 μm and 50nm, respectively. The CTC distances of double slits are 190 nm, 200 nm, 220 nm, 240 nm, 250 nm and 270 nm, respectively. Image Size: 2.4 μm×0.6 μm. Scale bars: 300 nm. **(b)** The line-scanning profiles of these simulated images by addressing the CTC distances of double slits.

By using the imaging theory, we provide a numerical simulation of images by using different microscopies. To simulate the imaging results from different types of microscopes, these double slits with the fixed 50-nm width and varied CTC distance are employed as the objects for keep the consistence with our experimental cases. To simulate the images of doubles slits by coherent bright-field microscope, the convolution between the objects and an Airy spot profile $\frac{J_1(krNA)}{krNA}$ with 0.9 *NA* is employed. Similarly, the simulation of traditional RSCM and objective-chip-based RSCM is implemented according to the theory of SCM, as mentioned in Ref [8]. Both PSFs of condenser lens and collector lens in the traditional RSCM are the Airy spots with the amplitude of $\frac{J_1(krNA)}{krNA}$, where NA=0.9 for a fair comparison. For our objective-chip-based RSCM, the PSF of condenser lens is the simulated electric field of our objective chip and the PSF of collector lens is Airy spot $\frac{J_1(krNA)}{krNA}$ with a measured *NA* of 0.83.

Supplementary Fig. 10 shows the simulated images that reveal the different resolution in various microscopies. As observed in our experiment, the coherent bright-field microscopy cannot resolve any double slit, suggesting its low resolving power. But, the traditional SCM by using refraction-based objective can only resolve the double slits with the CTC distance large than 240 nm, which agrees with our experimental results in Fig. 3 of main text.

To show the advantages of our proposed SCMs, we give a detailed comparison with the previously reported SCMs by presenting their various parameters. One can find that we have proposed the first planar-lens-based reflective SCM and the ultra-long working distance at the level of millimeter. More importantly, we can use industrial DUV lithography to fabricate all these objective chips in a mass-product and low-cost way, which pushes this objective chip towards practical applications. Supplementary Fig. 11 shows the images of our fabricated objective chips, which validates the feasibility for industrial mass production.

**Supplementary Table 1. A comparison among far-field label-free SCMs based on planar lenses**

| Reference | Material | Thickness of lens | Type of lens | λ | Ambient Medium | Working Mode | Working Distance | Substrate | Center To Center Distance |
|---|---|---|---|---|---|---|---|---|---|
| Rogers et al.[9] | Al | 100 nm | Amplitude | 640 nm | oil | Transmission | 10.3 μm | quartz | 315 nm(0.492λ) |
| Qin et al.[10] | Cr | 100 nm | Amplitude | 405 nm | air | Transmission | 55 μm | quartz | 228 nm(0.563λ) |
| Chen et al.[11] | TiO$_2$ | 600 nm | Phase | 532 nm | oil | Transmission | 125 μm | quartz | 400 nm(0.752λ) |
| Yuan et al.[12] | Gold | 100 nm | Amplitude | 800 nm | air | Transmission | 10 μm | quartz | 320 nm(0.4λ) |
| Wang et al.[13] | MoS$_2$ | 10 nm | Amplitude | 450 nm | air | Transmission | 20 μm | sapphire | 200 nm(0.444λ) |
| This work | SiO$_2$ | 527 nm | Phase | 405 nm | air | Reflection | 1000 μm | quartz | 200 nm(0.494λ) |

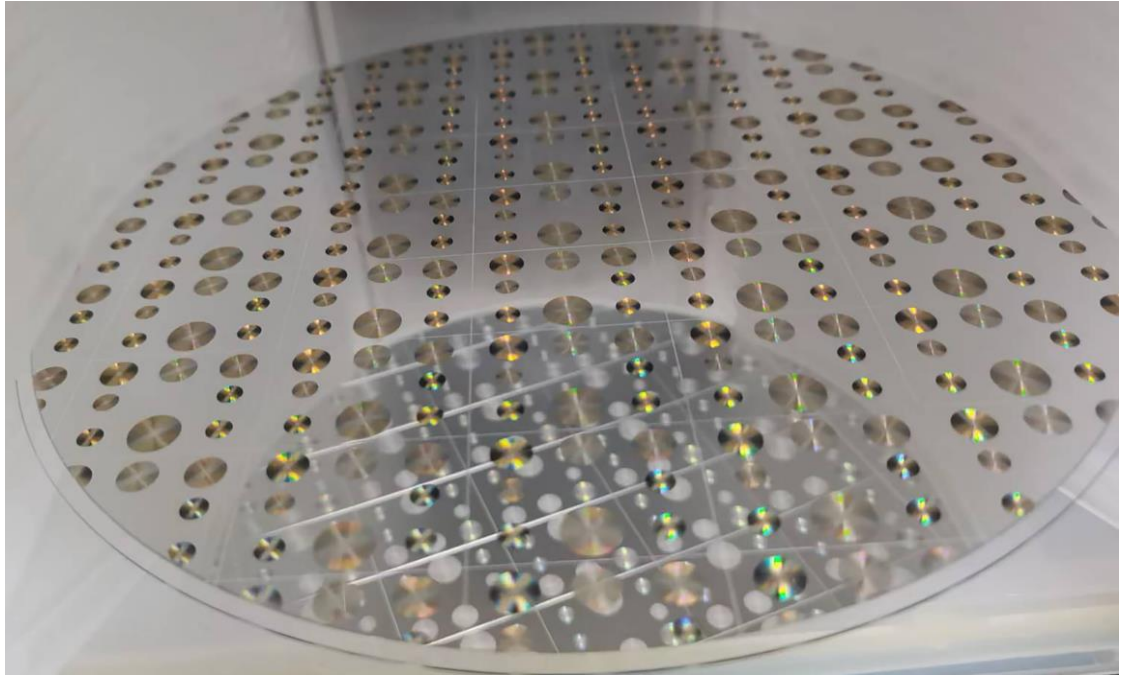

**Supplementary Figure 11. Fabricated objective chips on an 8-inch quartz wafer by using DUV lithography**

**References**

[1]. Li, M., Li, W., Li, H., Zhu, Y., & Yu, Y., Controllable design of super-oscillatory lenses with multiple sub-diffraction-limit foci, *Sci. Rep.* **7**, 1(2017).

[2]. Diao, J., Yuan, W., Yu, Y., Zhu, Y., & Wu, Y., Controllable design of super-oscillatory planar lenses for sub-diffraction-limit optical needles, *Opt. Express* **24**, 1924(2016).

[3]. Wan, X., Shen, B., & Menon, R., Diffractive lens design for optimized focusing, *J. Opt. Soc. Am. A* **31**, B27(2014).

[4]. Liu, T., Liu, J., Zhang, H., & Tan, J., Efficient optimization of super-oscillatory lens and transfer function analysis in confocal scanning microscopy, *Opt. Commun.* **319**, 31(2014).

[5]. Liu, T., Shen, T., Yang, S., & Jiang, Z., Subwavelength focusing by binary multi-annular plates: design theory and experiment, *J. Opt.* **17**, 035610(2015).

[6]. Zhang, Z., Li, Z., Lei, J., Wu, J., Zhang, K., Wang, S., Cao, Y., Qin, F., & Li, X., Environmentally robust immersion supercritical lens with an invariable sub-diffraction-limited focal spot, *Opt. Lett.* **46**, 2296(2021).

[7]. Fang, W., Lei, J., Zhang, P., Qin, F., Jiang, M., Zhu, X., Hu, D., Cao, Y., & Li, X., Multilevel phase supercritical lens fabricated by synergistic optical lithography, *Nanophotonics* **9**, 1469(2020).

[8]. Huang, K., Qin, F., Liu, H., Ye, H., Qiu, C. W., Hong, M., Luk'yanchuk, B., & Teng, J., Planar diffractive lenses: fundamentals, functionalities, and applications, *Advanced Materials* **30**, 1704556(2018).

[9]. Rogers, E. T., Lindberg, J., Roy, T., Savo, S., Chad, J. E., Dennis, M. R., & Zheludev, N. I., A super-oscillatory lens optical microscope for subwavelength imaging, *Nat Mater* **11**, 432(2012).

[10]. Qin, F., Huang, K., Wu, J., Teng, J., Qiu, C. W., & Hong, M., A supercritical lens optical label‑free microscopy: sub‑diffraction resolution and ultra‑long working distance, *Advanced Materials* **29**, 1602721(2017).

[11]. Chen, W. T., Zhu, A. Y., Khorasaninejad, M., Shi, Z., Sanjeev, V., & Capasso, F., Immersion Meta-Lenses at Visible Wavelengths for Nanoscale Imaging, *Nano Lett* **17**, 3188(2017).

[12]. Yuan, G., Rogers, K. S., Rogers, E. T., & Zheludev, N. I., Far-field superoscillatory metamaterial superlens, *Physical Review Applied* **11**, 064016(2019).

[13]. Wang, Z., Yuan, G., Yang, M., Chai, J., Steve Wu, Q. Y., Wang, T., Sebek, M., Wang, D., Wang, L., Wang, S., Chi, D., Adamo, G., Soci, C., Sun, H., Huang, K., & Teng, J., Exciton-Enabled Meta-Optics in Two-Dimensional Transition Metal Dichalcogenides, *Nano Letters*, 7964(2020).